\documentclass[aps,prd,showpacs,
superscriptaddress,nofootinbib]{revtex4-1}
\usepackage{tabularx}
\usepackage{multirow}
\usepackage{graphicx}
\usepackage{dcolumn}
\usepackage{float}
\usepackage{morefloats}
\usepackage{amsmath}
\usepackage{amssymb}
\usepackage{bm}
\usepackage{color}
\usepackage[normalem]{ulem}
\usepackage[dvipsnames]{xcolor}
\usepackage{hyperref}
\usepackage{placeins}

\hypersetup{
	colorlinks=true, % false: boxed links; true: colored links
	linkcolor=blue,  % color of internal links
	citecolor=cyan,  % color of links to bibliography
}

\DeclareMathAlphabet{\pazocal}{OMS}{zplm}{m}{n}

\begin{document}
	
\title[Trapping of null geodesics]
	{Trapping of null geodesics 
	in slowly rotating spacetimes}

\author{Jaroslav Vrba}
\email{jaroslav.vrba@physics.slu.cz}
\affiliation{Research Centre for Theoretical Physics and Astrophysics, 
	Institute of Physics in Opava, Silesian University in Opava, 
	Bezru\v{c}ovo n\'{a}m\v{e}st\'{i} 13, CZ-74601 Opava, Czech Republic}
	
\author{Martin Urbanec}
\email{martin.urbanec@physics.slu.cz}
\affiliation{Research Centre for Computational Physics and Data Processing, 
	Institute of Physics in Opava, Silesian University in Opava,
	Bezru\v{c}ovo n\'{a}m\v{e}st\'{i} 13, CZ-74601 Opava, Czech Republic}
	
\author{Zden\v{e}k Stuchl\'{i}k}
\email{zdenek.stuchlik@physics.slu.cz}
	
\affiliation{Research Centre for Theoretical Physics and Astrophysics, 
	Institute of Physics in Opava, Silesian University in Opava,
	Bezru\v{c}ovo n\'{a}m\v{e}st\'{i} 13, CZ-74601 Opava, Czech Republic}	
	
\author{John C. Miller}
\email{john.miller@physics.ox.ac.uk}
\affiliation{Department of Physics (Astrophysics), University of 
	Oxford, Keble Road, Oxford OX1 3RH, UK}

\date{\today}

\begin{abstract}
	
	Extremely compact objects containing a region of trapped null 
	geodesics could be of astrophysical relevance due to trapping 
	of neutrinos with consequent impact on cooling processes or 
	trapping of gravitational waves. These objects have previously 
	been studied under the assumption of spherical symmetry. In the 
	present paper, we consider a simple generalization by studying 
	trapping of null geodesics in the framework of the Hartle-Thorne 
	slow-rotation approximation taken to first order in the angular 
	velocity, and considering a uniform-density object with uniform 
	emissivity for the null geodesics. We calculate effective 
	potentials and escape cones for the null geodesics and how they 
	depend on the parameters of the spacetimes, and also calculate 
	the ``local'' and ``global'' coefficients of efficiency for the 
	trapping. We demonstrate that due to the rotation the trapping 
	efficiency is different for co-rotating and 
	retrograde null geodesics, and that trapping can occur even for 
	$R>3GM/c^2$, contrary to what happens in the absence of rotation.
\end{abstract}
%
%\vspace{2pc}
%
%\noindent{\it Keywords}: compact objects, rotation, trapped 
%	null geodesics, escape cones
%	\pacs{04.50.-h, 04.40.Dg, 97.60.Gb}
\maketitle
\section{Introduction}

Recently, there has been increasing interest in 
``ultra'' compact objects (ones which could mimic black holes), in 
connection with the detection of gravitational waves coming from 
mergers \cite{Barack19}. Also, a 
general correlation has been suggested between quasinormal modes and 
the parameters of unstable circular null geodesics \cite{Cardoso09}, 
although there could be some exceptions to this
\cite{Konoplya17,Toshmatov18,Toshmatov19a,Stuchlik19c}.

On the other hand, extremely compact objects are 
important also because of the possible existence of 
regions of trapped null geodesics that could be relevant for the trapping of 
gravitational waves \cite{Abramowicz97} or neutrinos \cite{Stuchlik09b}. 
The trapping region is always centered around a stable circular null 
geodesic whose existence represents a necessary condition for 
the existence of the trapping zone. 
Null geodesics inside extremely compact neutron stars 
may govern the motion of neutrinos, if the neutron stars are 
sufficiently cool \cite{Stuchlik09b}, and neutrino trapping can be 
important for several reasons: it can modify (decrease) the neutrino 
flow that reaches distant observers during the birth 
of a neutron star and just afterward; also, trapped neutrinos can 
substantially influence how it cools -- they could modify its internal 
structure due to induced internal flows, and could cause self-organizing 
of neutron star matter due to these flows, as discussed in the case of 
the internal Schwarzschild spacetimes \cite{Stuchlik09b}, 
or generalized internal Schwarzschild spacetimes 
modified by the presence of a cosmological constant 
\cite{Stuchlik12d}; for the relevance of the cosmological constant in 
astrophysical phenomena see \cite{Stuchlik20a}.

Models of extremely compact objects with trapping 
zones (so-called ``trapping compact objects'') were 
until now based on spherically symmetric spacetimes representing 
non-vacuum solutions of general relativity, or of an 
alternative gravity theory. The first detailed study of 
trapping spheres were made for internal Schwarzschild 
spacetimes, representing objects with uniform energy density 
\cite{Stuchlik09b}, for which it was explicitly demonstrated that they 
can include trapping spheres only if the radius (R) of the object 
concerned is smaller than $R=3GM/c^2=3r_g/2$, where M is its mass and 
$r_g$ is the related gravitational radius. Such objects must therefore 
be really extremely compact - their radius must be smaller than that of 
the (unstable) circular photon orbit (photosphere) of the external 
vacuum Schwarzschild spacetime. It has also been shown 
that the efficiency of the trapping increases monotonically with 
decreasing radius of the uniform sphere, down to $R = 9r_g/8$ (the 
minimum allowed for standard internal Schwarzschild spacetimes) 
\cite{Stuchlik09b}, for the special case related to 
gravastars see \cite{Konoplya19,Posada19}. Trapping of 
neutrinos for extremely compact objects in the braneworld scenario has 
been studied in \cite{Stuchlik11c}.

It is certainly interesting and relevant, to test the 
influence of rotation on the trapping of null 
geodesics. In the present paper, we approach this by 
making the simplest approximation for rotating compact objects of using 
the Hartle-Thorne slow-rotation approximation taken to first order in 
the angular velocity and applying it, once again, for constant-density 
objects. We thus use the standard internal Schwarzschild spacetime 
generated by the uniform energy density, but augmented 
with the off-diagonal term of the Lense-Thirring type arising from the 
Hartle-Thorne equation for the dragging of inertial frames 
\cite{Hartle68}. Such a solution represents just the 
simplest way to estimate the role of spacetime rotation in the trapping 
of null geodesics, but it demonstrates this in a 
clear and illustrative way. For estimating the 
efficiency of the trapping, we assume a uniformly 
distributed, isotropically and uniformly radiating source 
as in our previous work 
\cite{Stuchlik09b,Stuchlik12d}. Discussion of the 
applicability of null geodesics to neutrino motion in 
the interior of neutron stars can be found in \cite{Stuchlik09b}.

In section 2, we introduce the non-rotating internal Schwarzschild 
configuration with uniform energy density and the motion of trapped null 
geodesics connected to the existence of an internal 
spacetime with a stable sphere of null geodesics. We define the 
geometry, the effective potential of the motion along null geodesics, 
and the angles related to directions of emission from 
local sources needed for the construction of escape cones for null 
geodesics and their relation to trapping zones. In \mbox{section 3}, we 
introduce the first-order rotating Hartle-Thorne 
spacetime with uniform 
distribution of energy density and with the tetrad 
formalism being used for the observers (sources) in 
the rotating spacetime, and 
we study the null geodesics in both the internal and 
external first-order Hartle-Thorne spacetimes, 
constructing the related effective potentials for the null geodesic 
motion. In \mbox{section 4} we construct the escape 
cones (and complementary trapping cones) for the rotating and 
non-rotating configurations, and calculate their dependence on the 
compactness of the Hartle-Thorne object, its rotation, and their 
position in the object. In \mbox{section 5}, we define and calculate the 
corresponding \lq local \rq \ and \lq global\rq \ trapping efficiencies 
for null geodesics, and compare the trapping efficiencies of rotating 
and non-rotating configurations. In \mbox{section 6}, 
we discuss our results. Throughout, we use geometric units with $c = G = 
1$.

\section{Internal Schwarzschild spacetime and its null geodesics}

We here start by restricting attention to the simplest 
internal solution for our compact objects, with a uniform distribution 
of the energy density and with the internal geometry represented by the 
static and spherically symmetric internal Schwarzschild spacetime 
\cite{Schwarzschild1917}. We first present the 
geometry, its null geodesics, and the trapping of null geodesics 
occurring when the object is extremely compact.

\subsection{Internal Schwarzschild spacetime}

In standard Schwarzschild coordinates 
($t,r,\theta,\phi$), the line element of static and spherically symmetric 
spacetimes takes the form
\begin{equation}
\mathrm{d}s^2=-e^{2\Phi(r)}\mathrm{d}t^2+e^{2\Psi(r)}\mathrm{d}r^2 
+r^2\big(\mathrm{d}\theta^2 + \sin^2\theta \mathrm{d}\varphi^2\big).
\label{met}
\end{equation} 
The temporal and radial components of the internal Schwarzschild metric 
with the uniform energy density, $\rho=const$, are 
given by
\begin{equation}
(-g_{tt})^{1/2}=e^{\Phi}=\frac{3}{2}Y_1-\frac{1}{2}Y(r),\qquad 
(g_{rr})^{1/2}=e^{\Psi}=\frac{1}{Y(r)},
\end{equation}
where
\begin{eqnarray}
Y(r)=\Bigg(1-\frac{r^2}{a^2}\Bigg)^{1/2},\qquad 
Y_1=Y(R)=\Bigg(1-\frac{R^2}{a^2}\Bigg)^{1/2},\qquad
\frac{1}{a^2}=\frac{8}{3}\pi \rho=\frac{2M}{R^3},
\end{eqnarray}
with $R$ and $M$ being the total radius and mass of 
the object. At the surface $r=R$ the internal geometry is smoothly 
matched to the external vacuum Schwarzschild geometry
\begin{equation}
g_{tt}=-e^{2\Phi}=-\Bigg(1-\frac{2M}{r} \Bigg),\qquad 
g_{rr}=e^{2\Psi}=\Bigg(1-\frac{2M}{r} \Bigg)^{-1}.
\end{equation}
The radial profile of the pressure for the internal 
uniform density Schwarzschild spacetimes and its generalization to the 
case with non-zero cosmological constant is given and discussed in 
\cite{Stuchlik08a,Bohmer04}; the internal solutions for polytropic 
equations of state for the matter are discussed in 
\cite{Tooper64,Stuchlik16a,Novotny17,Stuchlik17} and their dynamical stability is discussed in 
\cite{Chandrasekhar1938,Hladik20,Posada20}.

In order to have a finite central pressure $P_\mathrm 
c = P(r=0)$, the surface radius has to fulfill the condition 
$R>R_\mathrm c = 9r_\mathrm g /8 \equiv 9M/4$; of course, 
at the surface $P(r=R)$ is always zero. Note that the 
internal Schwarzschild solutions with $R<R_\mathrm c$ are physically 
unrealistic, but they do have a physical meaning 
in a modified form with 
a very interesting interpretation for $R \to r_\mathrm g$, as they 
correspond in this limit to gravastars \cite{Ovalle19}. Here we consider 
the internal Schwarzschild solutions 
with $R>9r_\mathrm g /8$, focusing 
on those in the range $R \in (9/8, 1.6)r_\mathrm g$ corresponding to the 
so called trapping internal Schwarzschild spacetimes that contain 
regions of trapped null geodesics \cite{Stuchlik09b}.

\subsection{Effective potentials for the null geodesics}
The motion along null geodesics is governed by the relations
\begin{equation}
\frac{\mathrm{D}p^{\mu}}{\mathrm{d}\tau}=0,\qquad p_{\mu}p^{\mu}=0,
\label{geod}
\end{equation}
where $p^{\mu}$ is the 4-momentum and $\tau$ is the 
affine parameter. Two Killing vector fields ($\frac{\partial}{\partial 
t}\ \mathrm{and}\ \frac{\partial}{\partial \varphi}$) imply two conserved 
components of the 4-momentum:
\begin{equation}
p_t=-E\quad \mathrm{(energy)},\qquad p_{\varphi}=\phi\quad 
(\textrm{axial angular momentum})
\label{const}
\end{equation}
which are referred to as ``motion constants''. 
Motion in the spherically symmetric case is limited 
to a single plane. In the case of 
only one geodesic, it is convenient to choose 
that to be the equatorial plane of the coordinate 
system i.e. we set $\theta=\pi/2=$ const. Introducing the impact 
parameter $\lambda=\phi/E$ and using the 
normalisation condition, we obtain the relation
\begin{equation} 
(p^r)^2=\frac{-1}{g_{tt}g_{rr}}E^2\Bigg(1+g_{tt}\frac{\lambda^2}{r^2}\Bigg).
\end{equation}
It is obvious that the energy is not relevant 
(we can use it for scaling of the impact parameter $\lambda$). 
The expression in brackets has to be non-negative 
\cite{MTW}. We can then introduce the effective 
potential determining the turning points of the 
radial motion along the null geodesics for a given impact parameter 
$\lambda$.

\begin{equation}
\lambda^2\leq 
\mathrm{V}_\mathrm{eff}=\begin{cases}\mathrm{V}_\mathrm{eff}^\mathrm{int}
=\frac{4a^2[1-Y^2(r)]}{[3Y_1-Y(r)]^2}&\text{for } r \le R\\
\mathrm{V}_\mathrm{eff}^\mathrm{ext}=\frac{r^3}{r-2M}&\text{for } r>R
\end{cases}
\label{e:sveff}
\end{equation}

There is a local maximum of $V_\mathrm{eff}^\mathrm{int}$ corresponding 
to a stable circular null geodesic (in the 
interior), given by
\begin{eqnarray} 
r_\mathrm{c(i)}^2=\frac{R^3\left(\frac{4R}{9M}-1 \right)}{R-2M}, 
~~~\lambda_\mathrm{c(i)}^2=\frac{4a^2}{9Y_1^2-1}; 
\end{eqnarray} 
and a local minimum of $V_\mathrm{eff}^\mathrm{ext}$ 
corresponding to an unstable circular null geodesic 
(in the exterior), given by 
\begin{eqnarray} 
r_\mathrm{c(e)}=3M, 
~~~\lambda_\mathrm{c(e)}^2=27M^2. 
\end{eqnarray} 

\begin{figure}[h]
	\centering
	\includegraphics[width=0.60\hsize]{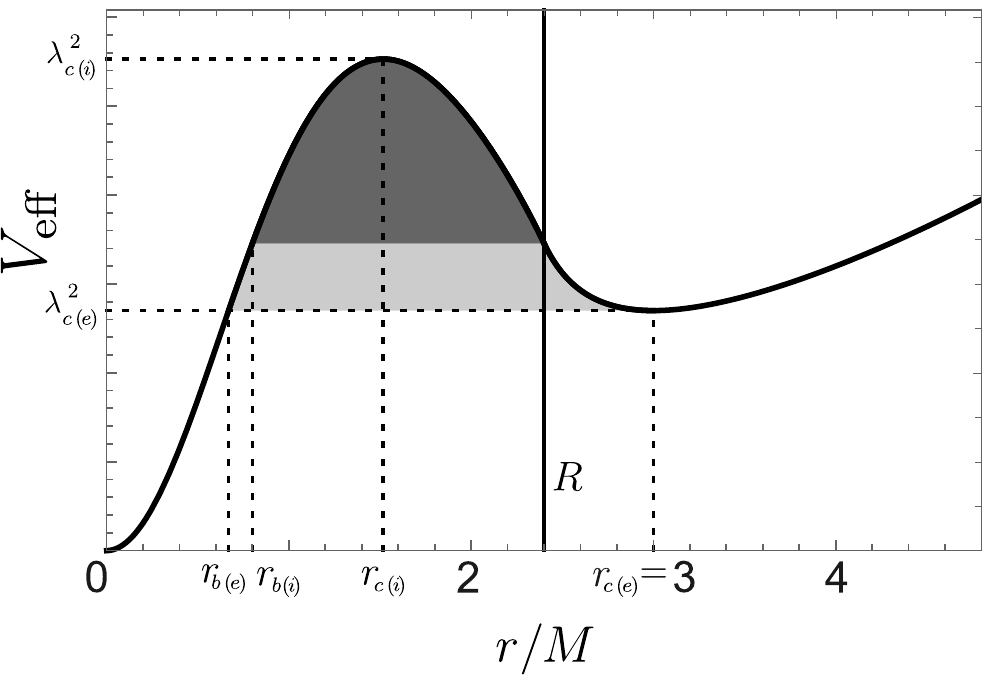}
	\caption{The effective potential for an object
		with $R=2.4M$; the trapped areas are shaded. The darkly-shaded 
		area, bounded by the radii $r_\mathrm{b(i)}$ and $R$, contains null 
		geodesics trapped completely inside the object, while the 
		lightly-shaded area, bounded by the radii $r_\mathrm{b(e)}$ and 
		$r_\mathrm{c(e)}$, contains null geodesics, which can temporarily pass 
		through the surface into the vacuum before returning.}
	\label{svef}
\end{figure}

Fig. \ref{svef} shows the effective potential of the trapping Schwarzschild 
spacetime with radius $R=2.4M$. The shaded areas 
indicate the zones of trapped null geodesics: the 
darker one corresponds to null geodesics fully contained in the 
internal spacetime (with inner boundary at $r_\mathrm{b(i)}$), while 
the lighter one is related to null geodesics that 
partially move in the external spacetime (with an inner boundary at 
$r_\mathrm{b(e)}$). The boundaries of the trapping regions 
lying within the internal Schwarzschild spacetime are given by (see 
\cite{Stuchlik09b}):

\begin{eqnarray}
r_\mathrm{b(i)}= \frac{R}{2R-3} \sqrt{ \frac{32R^2-144R+162}{2} }
\end{eqnarray}
and 
\begin{eqnarray}
\left(\frac{r_\mathrm{b(e)}}{R}\right)^2= &&\frac{ 
	27(10R^4-18R^3-108R+243) }{ 
	(2R^3+27)^2 }\\ \nonumber
&&-\frac{ 
	[6R^{3/2}(R-2)^{1/2}(R^4-108R+243)^{1/2}] }{ 
	(2R^3+27)^2 }.
\end{eqnarray}
	
For our analysis, it is important whether the 
effective potential is monotonic or not. If it is monotonic, then there 
cannot be any trapped area present. In other words, only if a local 
minimum (and a local maximum) of the effective potential exist, can 
there be a trapped area. For the 
internal Schwarzschild spacetimes, the analysis of the null geodesics 
shows that it is impossible to have regions of 
trapped null geodesics for $R>3M$.

Two types of trapped geodesic are distinguished in 
Fig. \ref{svef}. In the darkly-shaded 
part, the null geodesics are limited to the internal spacetime 
whereas in the lower lightly-shaded part, the null 
geodesics can pass through the surface of the object, 
although they are still bound and trapped 
(for details see \cite{Stuchlik09b}).

\subsection{Escape cones and trapped null geodesics}

In order to treat consistently the trapping of null geodesics in the 
context of neutrinos emitted by a matter of the 
trapping configuration, we need to determine the 
escape cones - the trapped neutrinos are then the 
proportion of the emitted ones not escaping.

We first briefly summarise the construction of the 
escape cones in the spherically symmetric internal uniform Schwarzschild 
spacetimes, as they are a starting point for our paper and provide a test 
for it. The escape cones need to be related to the 
static observers (sources) in the static spacetime and they are 
considered in the corresponding local frames of these 
static observers. The tetrad of the differential 
forms is defined as
\begin{eqnarray}
e^{(t)}=\mathrm e^\Phi \mathrm d t,~~ e^{(r)}=\mathrm e^\Psi \mathrm d 
r,~~e^{(\theta)}=r \mathrm d \theta,~~ e^{(\varphi)}=r 
\mathrm{sin}\theta \mathrm d \varphi,
\end{eqnarray}
and the line element then takes the special-relativistic form
\begin{equation}
\mathrm d s^2=-\left[ e^{(t)}\right]^2 + \left[ e^{(r)}\right]^2 + 
\left[ e^{(\theta)}\right]^2 + \left[ e^{(\varphi)}\right]^2.
\end{equation}

The complementary tetrad for the frame of the 
	4-vector $e_{(\alpha)}$ is given by
\begin{eqnarray}
e_{(\alpha)}^{\mu}e_\nu^{(\alpha)}=\delta^\mu_\nu,
~~e_\mu^{(\alpha)}e^\mu_{(\beta)}=\delta^\alpha_\beta.
\end{eqnarray}
Physically relevant projections of the 4-momentum 
$p^{\mu}$ are given by
\begin{eqnarray}
p^{(\alpha)}=p^{\mu}e_\mu^{(\alpha)},~~p_{(\alpha)}=p_\mu 
e^\mu_{(\alpha)}.
\end{eqnarray}

\begin{figure}[h]
	\centering
	\includegraphics[width=0.5\hsize]{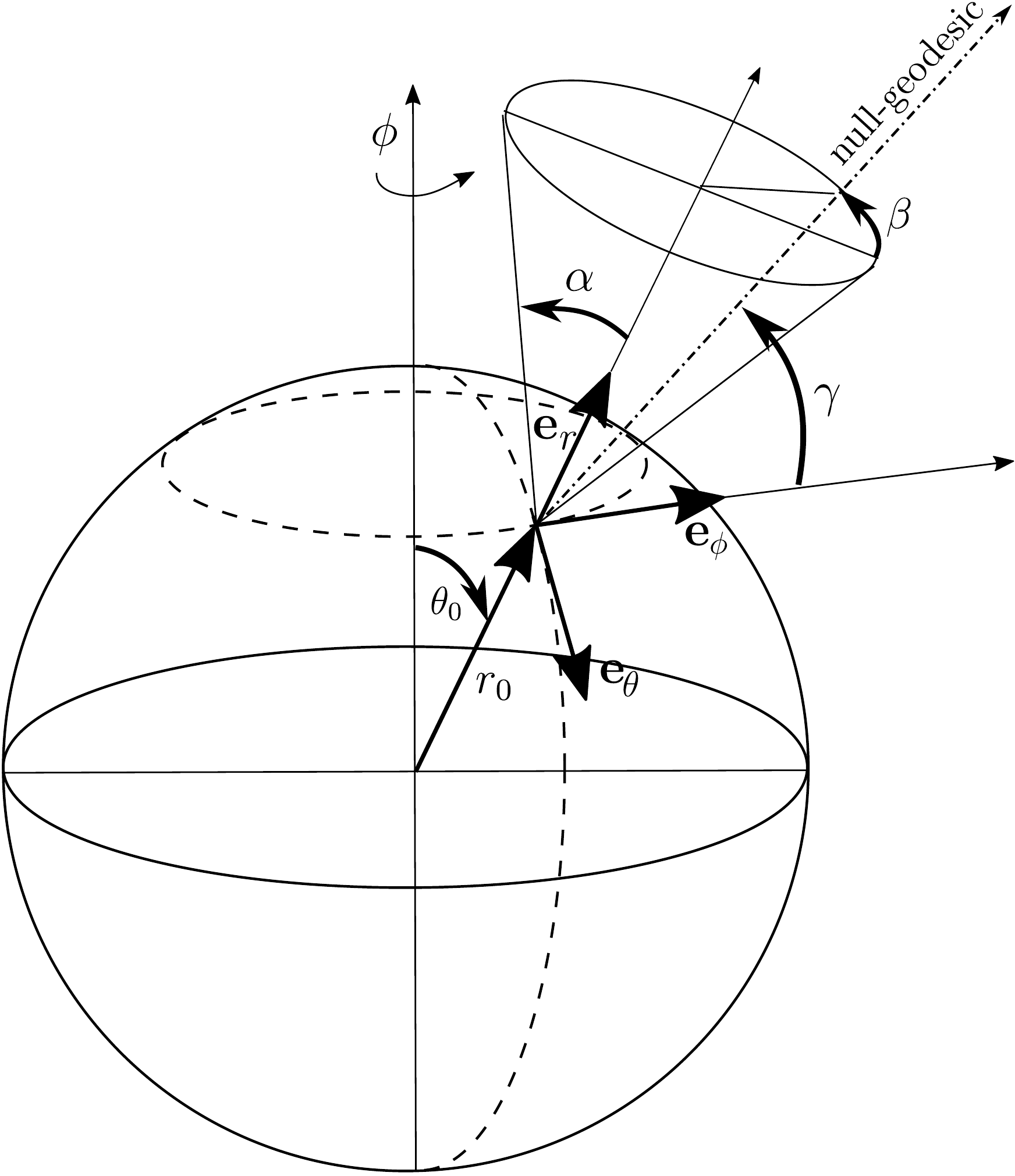}
	\caption{The definition of angles $\alpha$, $\beta$ and $\gamma$ 
		describing the direction of a null geodesic.}
	\label{f:defangles}
\end{figure}

The neutrinos radiated locally by a static source can be described by 
the directional angles, as demonstrated in Fig. \ref{f:defangles}. The 
angles {$\alpha,~\beta,~\gamma$} are connected by the relation
\begin{equation}
\cos\gamma=\sin\beta\sin\alpha,
\end{equation}
and are related to the outgoing radial direction in the spacetime.

The escape cone in the observer (source) frame is 
determined by the angles corresponding to photon parameters defining the 
stable and unstable null circular geodesics.

The angle $\alpha$ relates the directions of the null
geodesics and the outgoing radial unit vector. In the spherically 
symmetric internal Schwarzschild spacetime, a cone centered at the point 
of emission is determined by the angle $\alpha$, while the angle $\beta$ 
determines the position on the cone. 
We have to find the angles $\alpha_\mathrm{c(i)}$ 
corresponding to $\lambda_\mathrm{c(i)}$ of the stable null circular 
geodesic and $\lambda_\mathrm{c(e)}$ of the unstable null circular 
geodesic.

Now we have to relate the directional angles to the 
motion constants (and hence the impact parameters). 
Due to the spherical symmetry we can consider equatorial null geodesics 
(when $\beta=\pi/2$, or $\beta=3\pi/2$). The directional angle $\alpha$ is then 
governed by the relations ($p^{(\theta)}=0$)
\begin{equation}
\sin\alpha=\frac{p^{(\varphi)}}{p^{(t)}},
~~\cos\alpha=\frac{p^{(r)}}{p^{(t)}}.
\end{equation}

As the radial component of the 4-momentum is
\begin{equation}
p_r=\pm E \mathrm e^{\Psi-\Phi}\left(1-\mathrm 
e^{2\Phi}\frac{\lambda^2}{r^2}\right)^{1/2},
\end{equation}
we find the directional angle in the internal Schwarzschild spacetime in 
the form (putting for simplicity $M=1$)
\begin{eqnarray}
\sin\alpha&=&\left[\frac{3}{2}\left(1-\frac{2}{R}\right)^{1/2}
-\frac{1}{2}\left(1-\frac{2}{R}\left(\frac{r}{R}\right)^2\right)^{1/2}\right]
\frac{\lambda}{r}=A(r,R)\frac{\lambda}{r},\\
\cos\alpha&=&\pm\left(1-\sin^2\alpha\right)^{1/2}
\end{eqnarray}
To find the escape (trapping) cone in the region where trapping is 
possible, defined by the extension of the effective potential barrier, 
we have to calculate the angles for $\lambda=\lambda_\mathrm{c(i)}$ and 
$\lambda=\lambda_\mathrm{c(e)}$
\begin{eqnarray}
\sin\alpha_\mathrm{c(i)}(r,R)&=&A(r,R)\frac{R^{3/2}}{r(R-2)^{1/2}},\\
\sin\alpha_\mathrm{c(e)}(r,R)&=&A(r,R)\frac{3\sqrt{3}}{r}.
\end{eqnarray}
	
\begin{figure}[h]
	\centering
	\includegraphics[width=0.60\hsize]{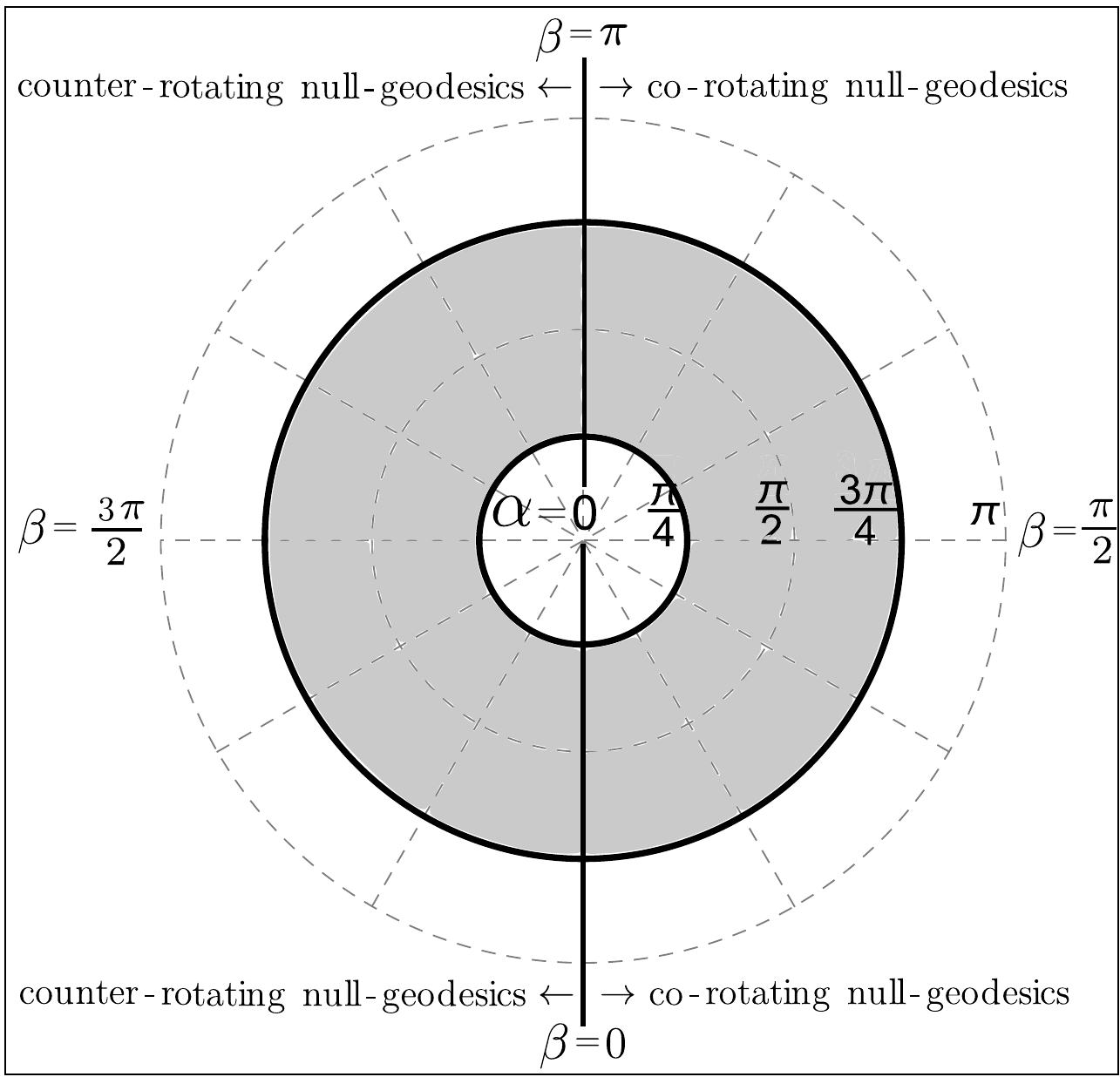}
	
	\caption{An example of an escape cone, for a 
	non-rotating configuration with $R/M=2.4$, located at the 
	position of the stable circular photon orbit. The figure 
	emphasizes the positions of the angle $\alpha$ = $ \{0, \pi / 4, 
	\pi / 2, 3 \pi / 4, \pi\}$, which has the character of a radial 
	coordinate in a classical polar graph and 
	the angle $\beta$ = $\{0, \pi / 2, \pi, 3 \pi / 2\}$, which has 
	the character of an angular coordinate in a classical polar 
	graph. The figure is also divided into 
	``left'' and ``right'' parts, for null 
	geodesics which are co-rotating 
	($\beta \in (0, \pi)$) and 
	counter-rotating ($\beta \in (\pi,2\pi)$).}
\label{scone}
\end{figure}

One always has 
$\alpha_\mathrm{c(i)}>\alpha_\mathrm{c(e)}$, and the trapping zone lies 
between these angles, as shown in Fig. \ref{scone} where 
the trapping zone is shaded. The escape cone (zone) 
is unshaded and null geodesics there can escape to infinity even if 
originally emitted inwards \footnote{Note that in the 
rotating spacetimes the symmetry of the trapping zone (cone) is lost as 
the motion depends on the sign of the impact parameter, as demonstrated 
for the case of trapped null geodesics in Kerr spacetimes 
\cite{Stuchlik10}}.

The escape and complementary trapping cones have 
been applied to calculate the trapping efficiency for 
null geodesics in 
spherically-symmetric uniform-density internal Schwarzschild spacetimes, 
and to estimate the neutrino trapping \cite{Stuchlik09b}. Here we 
generalize those results to the simplest case of 
first-order Hartle-Thorne models based on 
spherically-symmetric uniform-density internal Schwarzschild spacetimes 
as the background solution, augmented by the first order 
rotational 
off-diagonal metric coefficient. The first-order 
Hartle-Thorne model is 
equivalent to the Lense-Thirring metric in the 
exterior of the compact object, but has a different interior spacetime 
with regard to both the matter and rotational contributions.

\FloatBarrier
\section{First-order Hartle--Thorne spacetime with 
uniformly distributed energy density}

As already mentioned, the effects of rotation on 
the trapping internal Schwarzschild spacetime with uniformly distributed 
energy density is here being treated in the framework of 
the Hartle--Thorne slow-rotation model taken to first 
order in the angular velocity. The 
Hartle--Thorne models treat both the interior and exterior of 
relativistic perfect-fluid, rotating objects by expanding up to second 
order in the angular velocity $\Omega$ as measured by distant static 
observers, with $\Omega$ taken to be constant throughout the model 
\cite{Hartle68}. The case with the uniform 
internal Schwarzschild spacetime being taken as the reference model 
about which to perturb, has previously been studied by Chandrasekhar 
and Miller \cite{Chandrasekhar74}. For the present purposes, we are 
including only the first-order term here. We start by summarising the 
results of \cite{Hartle68} that will be applied here.

\subsection{Hartle--Thorne spacetimes and their locally non-rotating frames}

Using the frame formalism, the internal Hartle--Thorne metric can be 
written using the 1-forms 
\begin{equation}
\mathrm d s^2 = -\left[\mathrm e^{(t)} \right]^2 + \left[\mathrm e^{(r)} 
\right]^2 + \left[\mathrm e^{(\theta)} \right]^2 +\left[\mathrm 
e^{(\varphi)} \right]^2
\end{equation}
where the 1-forms of the metric can be conveniently defined to reflect 
the so-called locally non-rotating frames (LNRFs) 
\cite{Bardeen72}, corresponding to observers with 
zero angular momentum at given $(r,\theta)$ (therefore, the term ZAMO 
frames is also used) who are locally co-rotating 
with respect to the internal Hartle-Thorne spacetime, being dragged to 
corotation with the angular velocity $\omega(r)$ characterizing the 
spacetime rotation relative to static distant observers:
\begin{eqnarray}
&&\mathrm e^{(t)}=\sqrt{-g_{tt(\mathrm i)}} \mathrm d t, 
~~\mathrm e^{(r)}=\sqrt{g_{rr(\mathrm i)}} \mathrm d r, \\ \nonumber
&&\mathrm e^{(\theta)}=\sqrt{g_{\theta\theta(\mathrm i)}} 
\mathrm d \theta, ~~
\mathrm e^{(\varphi)}=\sqrt{g_{\theta\theta(\mathrm i)}}\sin\theta
(\mathrm d \varphi - \omega \mathrm d t).  
\end{eqnarray}
The metric coefficients are given in 
\cite{Hartle68}; $\omega(r)$ is the angular 
velocity of the LNRFs reflecting the rotational dragging of the internal 
spacetime by the gravitation of the object rotating with the angular 
velocity $\Omega= const$. The motion of the particles constituting 
the Hartle--Thorne object corresponds to rigid and 
uniform rotation with angular velocity $\Omega$ relative to distant 
static observers, so that their 4-velocity $u^\mu$ 
has components
\begin{eqnarray}
u^t=\left(-g_{tt} - 2\Omega g_{t\varphi} -g_{\varphi\varphi}\Omega^2 
\right)^{-1/2}, ~~~
u^\varphi=\Omega u^t, u^r=u^\theta=0 . 
\end{eqnarray}

The angular velocity of the rotating matter relative 
	to the rotating spacetime
\begin{equation}
\bar\omega(r) \equiv \Omega-\omega(r)
\end{equation}
enters the Einstein gravitational equations -- it is of 
first order in $\Omega$ and satisfies the 
differential equation
\begin{equation}
\frac{1}{4}\frac{\mathrm{d}}{\mathrm{d}r}
\Bigg(r^4f(r)\frac{\mathrm{d}\bar{\omega}}{\mathrm{d}r}\Bigg)
+\frac{4}{r}\frac{\mathrm{d}f(r)}{\mathrm{d}r}\bar{\omega}=0,
\label{difo}
\end{equation}
where 
\begin{equation}
f(r)=\Bigg(\frac{-1}{g_{tt(i)}g_{rr(i)}}\Bigg)^{1/2}.
\label{e:f}
\end{equation}

The solution of this equation needs to be regular at 
the origin and so we have $\frac{d\bar\omega}{dr}=0$ at $r=0$ while 
$\bar\omega$ goes to a finite value $\bar\omega_c$ there which is found 
by matching with the vacuum exterior solution at the surface $r=R$. That 
matching gives expressions for the angular momentum $J$ and angular 
velocity $\Omega$ of the object \cite{Hartle68}
\begin{equation}
J=\frac{1}{6}R^4\Bigg(\frac{\mathrm{d}\bar{\omega}}{\mathrm{d}r} 
\Bigg)_{r=R},
\label{surface1}
\end{equation}
and 
\begin{equation}
\Omega=\bar\omega(R) + \frac{2J}{R^3}.
\label{surface2}
\end{equation}
o complete the solution, it is convenient to first 
solve Eq. (\ref{difo}) for $\bar\omega$ measured in units of 
$\bar\omega_c$, then to use (\ref{surface1}) and (\ref{surface2}) to find 
$J/\bar\omega_c$ and $\Omega/\bar\omega_c$, and finally to use these for 
obtaining $J$ and $\bar\omega(r)$ measured in units of $\Omega$. The 
first of these gives the moment of inertia $I=J/\Omega$ and is related 
to the radius of gyration by (see \cite{Abramowicz93}):
\begin{equation} 
\pazocal R_\mathrm{gyr}=\left( \frac{J}{\Omega M} \right)^{1/2}. 
\end{equation}

The external Hartle--Thorne geometry can be described in the tetrad 
frame taking the form \cite{Hartle68}
\begin{eqnarray}
&&\mathrm e^{(t)}=\sqrt{-g_{tt(\mathrm e)}} \mathrm d t, 
~~\mathrm e^{(r)}=\sqrt{g_{rr(\mathrm e)}} \mathrm d r, \nonumber\\ 
&&\mathrm e^{(\theta)}=\sqrt{g_{\theta\theta(\mathrm e)}} \mathrm d \theta, 
~~ \mathrm e^{(\varphi)}=\sqrt{g_{\theta\theta(\mathrm e)}}\sin\theta
(\mathrm d \varphi - \frac{2J}{r^3} \mathrm d t).
\end{eqnarray}
The form of the metric coefficients can be found in 
\cite{Hartle68} or, in an alternative form, in 
\cite{Abramowicz03}.

\subsection{First-order Hartle--Thorne spacetime}

The line element of the first-order Hartle--Thorne internal 
spacetime, with uniformly distributed energy 
	density, considered in the following study takes the form
\begin{equation}
\mathrm{d}s^2=-e^{2\Phi(r)}\mathrm{d}t^2+e^{2\Psi(r)}\mathrm{d}r^2 
+r^2\big(\mathrm{d}\theta^2 + \sin^2\theta \mathrm{d}\varphi^2\big) 
- 2 \omega(r) r^2 \sin^2\theta \mathrm{d}t\mathrm{d}\varphi,
\label{metLTi}
\end{equation} 
where $\Phi(r)$ and $\Psi(r)$ are the functions governing the static 
spherically symmetric configuration, here taken to 
	be the internal Schwarzschild spacetime, and $\omega(r)$ is the angular 
velocity of the LNRF related to those spacetimes. The rotating spacetime 
geometry is then determined by finding the solution of Eq. (\ref{difo}) for 
the angular velocity of the rotating matter relative to the spacetime 
geometry, $\bar\omega$, where the characteristic function of the 
solution is
\begin{equation}
f(r)=\Bigg(\frac{4Y^{2}(r)}{(3Y_{1}-Y(r))^{2}}\Bigg)^{1/2}.
\end{equation}

\begin{figure}[h]
	\centering
	\includegraphics[width=0.55\hsize]{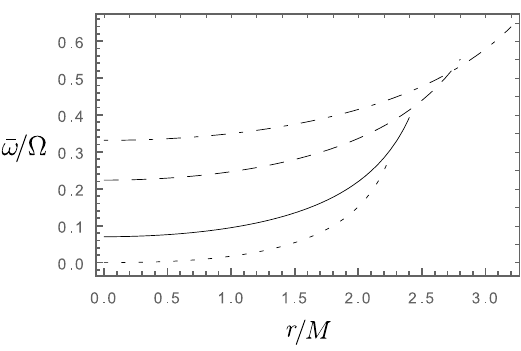}
	\caption{Numerical solution for $\bar{\omega}$
		scaled by $\Omega$ for $R/M=2.25$ (dotted line), 
		$R/M=2.4$ (solid line), $R/M=2.8$ (dashed line) and 
		$R/M=3.2$ (dash-dotted line).}
	\label{f:omega}
\end{figure}

The results of the integration are presented in Fig. \ref{f:omega} -- we 
can see how the frame dragging increases with increasing compactness, if 
the angular velocity $\Omega$ is kept fixed. The strongest dragging 
effect, with $\omega(r=0)=\Omega$ and $\bar{\omega}(r=0)=0$, is obtained 
for the maximally compact case with $R/M=2.25$. Note 
that this represents a limiting case, with the central pressure going to 
infinity, which cannot itself be considered as at all physical. In the 
following, we will focus on the three other cases shown here as being 
representative of these very compact models.

The metric of the external first-order Hartle-Thorne 
(or Lense-Thirring) 
spacetime takes the form
\begin{eqnarray}
\mathrm{d} s^2 = && -\left( 1-\frac{2M}{r} \right) \mathrm{d}t^2+\left( 
1-\frac{2M}{r} \right)^{-1}\mathrm{d}r^2 +r^2\left(\mathrm{d}\theta^2 + 
\sin^2\theta \mathrm{d}\varphi^2\right)\nonumber \\
&& - 2 \frac{J}{r} \sin^2\theta 
\mathrm{d}t\mathrm{d}\varphi,
\label{metLTe}
\end{eqnarray} 

It is useful to consider also the inverse form of 
the metric representing the first-order spacetimes. 
The contravariant 
form of the internal first-order Hartle-Thorne metric 
is given by
\begin{eqnarray}
	\left(\frac{\partial}{\partial s}\right)^{2}_{\mathrm{int}} = &&-e^{-2\Phi(r)}_\mathrm{int} \frac{\partial^2}{\mathrm{\partial}t^2}+\left(  1-\frac{2 r^2}{R^3}   \right) \frac{\partial^2}{\mathrm{\partial}r^2} + frac{1}{r^2}\left(\frac{\partial^2}{\mathrm{\partial}\theta^2} + \frac{1}{\sin^2\theta} \frac{\partial^2}{\mathrm{\partial}\varphi^2}\right)\nonumber \\
&&-\omega(r)e^{-2\Phi(r)}_\mathrm{int} \frac{\partial^2}{\mathrm{\partial}t\mathrm{\partial}\varphi},
\end{eqnarray}
where
\begin{equation}
e^{-2\Phi(r)}_\mathrm{int}= \left( \frac{2}{\sqrt{1-\frac{2 r^2}{R^3}}-3 
	\sqrt{\frac{R-2}{R}}} \right)^2 ,
\end{equation}
while the external first-order metric is 
\begin{eqnarray}
	\left(\frac{\partial}{\partial s}\right)^{2}_{\mathrm{ext}} =&& -\left(  1-\frac{2M}{r}   \right)^{-1}  \frac{\partial^2}{\mathrm{\partial}t^2}+\left(  1-\frac{2M}{r}   \right) \frac{\partial^2}{\mathrm{\partial}r^2} \frac{1}{r^2}\left(\frac{\partial^2}{\mathrm{\partial}\theta^2} + \frac{1}{\sin^2\theta} \frac{\partial^2}{\mathrm{\partial}\varphi^2}\right)\nonumber \\
&& -\frac{2 J}{(r-2) r^2} \mathrm{d}t\mathrm{d}\varphi.
\end{eqnarray}

It is very important that the $g^{t\phi}$ metric 
coefficient appears in a form independent of the 
latitudinal coordinate $\theta$ due to the applied 
first-order approximation. 
This fact enables separability of the equations for geodesic motion in the 
first-order Hartle-Thorne spacetimes of the type 
being considered here and substantially simplifies 
the discussion of the trapping effect of null geodesics, as we shall see 
later.

In the discussion of trapping, we use the fact that 
the ZAMO observers of the first-order Hartle-Thorne 
internal spacetime 
have the simple form of four velocity
\begin{equation}
( U^{\mu})_{\mathrm{ZAMO}} = e^{-\Phi(r)}(1,0,0,\omega), 
(U_{\mu})_{\mathrm{ZAMO}} = e^{\Phi(r)}(-1,0,0,0) .
\end{equation}

Similarly, the LNRF tetrad of 1-forms takes the 
simple form
\begin{eqnarray}
&&e^{(t)}_{\mu} = e^{\Phi(r)}(1,0,0,0),\nonumber\\  
&&e^{(r)}_{\mu} = e^{\Psi(r)}(0,1,0,0),  \nonumber\\
&&e^{(\theta)}_{\mu} = r(0,0,1,0), \nonumber\\
&&e^{(\varphi)}_{\mu} = r\sin\theta(-\omega,0,0,1), 
\end{eqnarray}
while the tetrad for a comoving frame in 1-forms 
also takes a simplified form
\begin{eqnarray}
&&\bar{e}^{(t)}_{\mu}= e^{\Phi(r)}(1,0,0,\bar{\omega}),\nonumber\\
&&\bar{e}^{(r)}_{\mu}= e^{\Psi(r)}(0,1,0,0),\nonumber\\
&&\bar{e}^{(\theta)}_{\mu}= r(0,0,1,0),\nonumber\\
&&\bar{e}^{(\varphi)}_{\mu}= r\sin\theta(-\Omega,0,0,1).
\end{eqnarray}

\subsection{Null geodesics of first-order Hartle--Thorne 
spacetime}
The Lagrangian governing test particle motion along 
	geodesics in both the internal and external 
first-order Hartle-Thorne 
spacetimes can be written as \cite{Chandrasekhar98} \footnote{We 
	follow only the procedure used there and not the 
		signature of the metric tensor.}
\begin{equation}
2\mathcal{L}=g_{tt}\dot{t}^2+g_{rr}\dot{r}^2 
+g_{\theta\theta}\dot{\theta}^2+g_{\varphi\varphi}\dot{\varphi}^2
+2g_{t\varphi}\dot{\varphi}\dot{t}.
\end{equation}
The metric functions (\ref{metLTi}) and 
(\ref{metLTe}) depend only on the radial and 
latitudinal coordinates, and so there are two 
Killing vector fields (time and axial) implying the 
	existence of two constants of motion
\begin{eqnarray}
&&p_t=\frac{\partial \mathcal{L}}{\partial\dot{t}} 
=g_{tt}\dot{t}+g_{t\varphi}\dot{\varphi}=-E=\mathrm{constant},\nonumber\\
&&p_{\varphi}=\frac{\partial\mathcal{L}}{\partial\dot{\varphi}} 
=g_{t\varphi}\dot{t}+g_{\varphi\varphi}\dot{\varphi}=\phi 
=\mathrm{constant}.
\label{e:e:constEL}
\end{eqnarray}
These constants correspond to the energy and the 
axial component of the angular momentum of test particles, as measured 
by static observers at infinity.

The geodesic motion is governed by the Hamilton-Jacobi equation related 
to the metric \mbox{tensor $g^{\mu\nu}$}
\begin{equation}
2\frac{\partial S}{\partial \tau}=g^{\mu\nu}\frac{\partial S}{\partial 
	x^{\mu}}\frac{\partial S}{\partial x^{\nu}},
\label{HJ}
\end{equation}
where $S$ denotes Hamilton's principal function, the so-called Action 
function. Assuming separability of variables, as enabled by 
the fact that $g^{t\phi}$ is independent of $\theta$, 
as mentioned above, we seek the principal function in the form
\begin{equation}
S=\frac{1}{2}\delta\tau - Et + \phi\varphi + S_r(r) + 
S_{\theta}(\theta),
\label{action}
\end{equation}
where $p^{\mu}p_{\mu}=\delta$. After a little algebra and introducing a 
separation constant $L$ serving as a motion constant related to the 
total angular momentum, the components of the 
particle four momentum can be expressed as follows
\begin{eqnarray}
&&p^r=g^{rr}p_r=g^{rr}\frac{\mathrm{d}S_r}{\mathrm{d}r}
=\Big(2g^{t\varphi}g^{rr}E\phi+g^{rr}\delta-g^{tt}g^{rr}E^2
-Lg^{rr}g^{\theta\theta} \Big)^{1/2},
\label{e:pr}\\
&&p^{\theta}=g^{\theta\theta}p_{\theta}
=g^{\theta\theta}\frac{\mathrm{d}S_{\theta}}{\mathrm{d}\theta}
=g^{\theta\theta}\Bigg(L-\frac{\phi^2}{\sin^2\theta} \Bigg)^{1/2},
\label{e:pth}\\
&&p^{\varphi}=\frac{\phi g_{tt}+Eg_{t\varphi}}
{g_{tt}g_{\varphi\varphi}-g_{t\varphi}^2},\\
&&p^{t}=\frac{\phi g_{t\varphi}+Eg_{\varphi\varphi}}
{g_{t\varphi}^2-g_{tt}g_{\varphi\varphi}}.
\end{eqnarray}
Motion along null geodesics corresponds to 
$\delta=0$. From now on we consider the components 
of the four-momentum as components of wave-vectors 
$k^{\mu}$ and $k_{\mu}$, as usual for null geodesics. 
The null geodesics (neutrino trajectories) are 
not dependent on the energy and so we can rescale 
the motion equations in terms of the energy to get 
new motion constants (impact parameters):
\begin{eqnarray}
\lambda=\frac{\phi}{E}\quad \mathrm{and}\quad \pazocal{L}=\frac{L}{E^2}.
\label{e:constlal}
\end{eqnarray}
The regions allowed for motion in the radial and 
latitudinal directions ($r-\theta$) are governed by the turning points, 
where $k^{r}=0$ or $k^{\theta}=0$. Using 
(\ref{e:constlal}) in (\ref{e:pr}) and (\ref{e:pth}), we arive 
at the effective potentials governing the radial 
and latitudinal motion in the form
\begin{equation}
\pazocal{L}_{r}=2g^{t\varphi}g_{\theta\theta}\lambda
-g^{tt}g_{\theta\theta},
\label{e:lr}
\end{equation}
\begin{equation}
\pazocal{L}_{r}(r,\lambda)=\begin{cases}
\frac{4 r^2\left(1-2\omega\lambda \right)}
{\left(\sqrt{1-\frac{2 r^2}{R^3}}-3 \sqrt{\frac{R-2}{R}}\right)^2}
&\text{for } r \le R\\
\frac{r^3-4J\lambda}{r-2} &\text{for }  r>R
\end{cases}
\label{e:lr2}
\end{equation}
and
\begin{equation}
\pazocal{L}_{\theta}(\theta,\lambda)=\frac{\lambda^2}{\sin^2\theta}.
\label{fcelth}
\end{equation}
Note that the latitudinal effective potential is 
independent of the spacetime parameters, whereas 
the radial effective potential does depend on them, 
namely on the parameter $J$ of the external spacetime, and 
on the parameter $R/M$ and the angular velocity 
function $\bar{\omega}(r)$ of the internal spacetime.

The reality conditions $k^{r} \geq 0$ and $k^{\theta} \geq 0$ imply 
restrictions on the impact parameter $\pazocal{L}$ in the form
\begin{equation}
\pazocal{L}_r(r,\lambda) \geq \pazocal{L} \geq \pazocal{L}_{\theta}(\theta,\lambda)
\end{equation}

For the behavior of the null geodesics, the character of the effective 
potential $\pazocal{L}_{r}(r,\lambda)$ is crucial as its extremal points 
determine the circular null geodesics giving basic restrictions on the 
trapping effect. The local extrema of the effective potential 
$\pazocal{L}_{r}(r,\lambda)$ are given by the condition 
$d\pazocal{L}_{r}/dr=0$ that is fulfilled where 
$\lambda=\lambda_{c}(r)$. Using the relation (\ref{e:lr2}), we arrive at 
the general formula
\begin{equation}
\lambda_{c}(r)=\frac{\frac{\mathrm{d}}{\mathrm{d}r}(g^{tt}g_{\theta\theta})}
{\frac{\mathrm{d}}{\mathrm{d}r}(2g^{t\varphi}g_{\theta\theta})}.
\label{lamc}
\end{equation}
The function $\lambda_c(r)$ determines the locations of 
the stable circular null geodesics in the internal spacetime (at 
$r_\mathrm{c(i)}$), and of the unstable ones in the exterior (at 
$r_\mathrm{c(e)}$). In the external spacetime, this formula corresponds 
to minima (if they exist) of the radial effective potential and 
takes the form
\begin{equation}
\lambda_{c}(r,J)=-\frac{(r-3) r^2}{2J} , 
\end{equation}
while in the internal spacetime it corresponds to 
the maxima (if they exist) and has the form
\begin{equation}
\lambda_{c}(r;R,\omega)=\frac{R^3 h}{r \left(h R^3 +2 r^2\right) \frac{\mathrm{d}\omega(r)}{\mathrm{d}r}+R^3 2 \omega(r) h},
\end{equation}
where
\begin{equation}
h(r;R)=3 \sqrt{\frac{R-2}{R}} \sqrt{1-\frac{2 r^2}{R^3}}-1.
\end{equation}

The fundamental limiting condition on the relevance of the local extrema 
of the radial effective potential is given by the relation 
$\pazocal{L}_{r} \geq \pazocal{L}_{\theta}$. We thus obtain the limiting 
functions $\lambda_{r\pm}(r,\theta)$, given by the condition 
$\pazocal{L}_{r}=\pazocal{L}_{\theta}$, in the general form
\begin{equation}
\lambda_{r\pm}=\sin^2{\theta}\Big( g^{t\varphi}g_{\theta\theta} \pm 
\sqrt{(g^{t\varphi}g_{\theta\theta})^2-g^{tt}g_{\theta\theta}\csc^2\theta} 
\Big),
\label{lamlim}
\end{equation}
that, for the internal spacetime, takes the form 
\begin{equation}
\lambda_{r\pm}^\mathrm{int}= \sin ^2\theta\left( -\frac{4 r^2 \omega(r) 
}{\left(\sqrt{1-\frac{2 r^2}{R^3}}-3 \sqrt{\frac{R-2}{R}}\right)^2}\pm 2 
\sqrt{\frac{r^2 \csc ^2\theta}{\left(\sqrt{1-\frac{2 r^2}{R^3}}-3 
		\sqrt{\frac{R-2}{R}}\right)^2}}\right),
\end{equation}
and, for the external spacetime, takes the form 
\begin{equation}
\lambda_{r\pm}^\mathrm{ext}=\sin ^2\theta \left(-\frac{2 J}{r-2}\pm 
\sqrt{\frac{r^3 \csc^2\theta}{r-2}} \right).
\end{equation}

The radial effective function $\pazocal{L}_r$ is thus physically 
relevant in the regions determined by the condition
\begin{equation}
\lambda_{r-}(r,\theta) \leq \lambda \leq \lambda_{r+}(r,\theta). 
\end{equation}								

\begin{figure}[ht]
	\centering
	\includegraphics[width=1.\hsize]{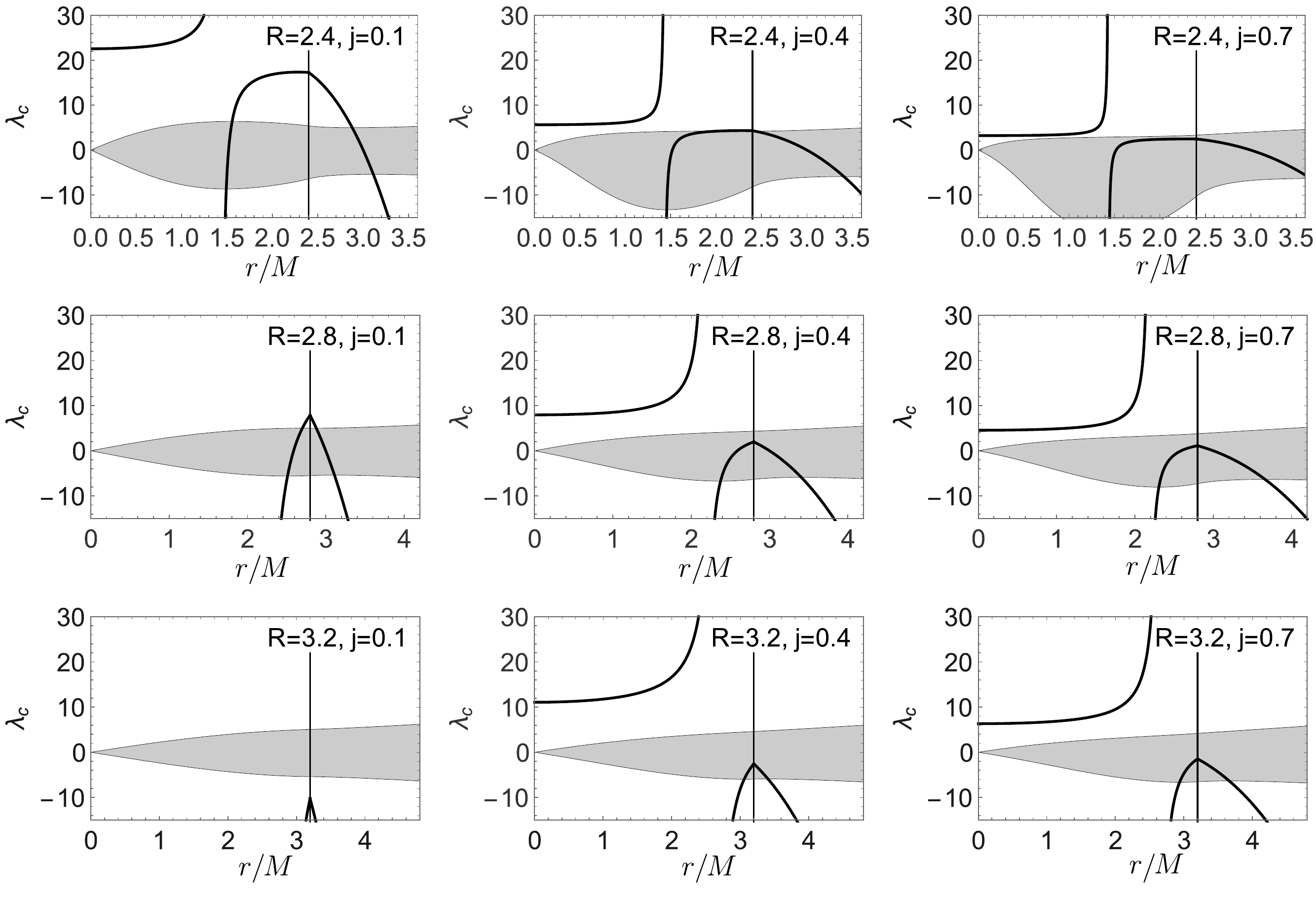}
	\caption{Restriction for the values of 
			$\lambda$ and the function for $\lambda_c$ in the equatorial 
			plane. The grey area indicates the allowed region for $\lambda$ 
			with the boundaries given by $\lambda_+$ and $\lambda_-$. The 
			bold solid line shows the values of $\lambda_c$ which marks the 
			extremes of $\pazocal{L}$. The first column is for $j=0.1$, the 
			second is for $j=0.4$ and the last is for $j=0.7$. The first row 
			is for $R/M=2.4$, the second is for $R/M=2.8$ and the last is 
			for $R/M=3.2$.}
	\label{f:lclthj}
\end{figure}

In the following discussion we usually consider the 
most extended region corresponding to the equatorial plane where 
$\sin\theta=1$. In Fig. \ref{f:lclthj} we demonstrate the situation 
corresponding to the trapping of the null geodesics, giving both the 
functions $\lambda_{r\pm}(r,\theta=\pi/2)$ and $\lambda_{c}(r)$ for a 
few characteristic values of the spacetime parameters, namely for 
different rotation rates given by $j$ ($j=J/M^2$ - the dimensionless 
angular momentum) and for our three highlighted values 
of the inverse compactness of the object $R/M$ (the 
compactness being defined as $C=M/R$, with $R$ being measured here in 
the equatorial plane). The regions allowed for the motion with a fixed 
value of the impact parameter $\lambda$ are located between the curves 
given by Eq. (\ref{lamlim}) (grey area). The location of the extremal 
points of the radial effective potential, corresponding to the circular 
null geodesics, is given by the function $\lambda_{c}(r)$.

Note that the stable circular photon orbit for 
the most highly compact object with $R/M=2.4$ is 
situated in an almost constant position for most 
allowed values of $\lambda_c$, while the unstable 
circular photon orbit changes its position more significantly 
for allowed $\lambda_c$. We 
	also point out the non-existence of circular photon orbits for low 
rotations and the less compact objects with 
$R/M=3.2$ -- this is in line with expectations, because for non-rotating 
configurations with $R/M>3$, there are no circular 
	null geodesics.

\subsection{Effective potential of the radial motion}

Now we discuss motion along the null geodesics in the radial direction 
demonstrating the behavior of the radial effective potential 
$\pazocal{L}_{r}(r)$ given by Eq. (\ref{e:lr}). In this 
subsection, we will use the terms $V_{eff}$ and $\pazocal{L}_{r}$ 
interchangeably. If the rotation is zero ($g^{t\varphi}=0$), Eq. 
(\ref{e:lr}) matches (\ref{e:sveff}), the effective potential for the 
internal (and external) Schwarzschild spacetime.

\begin{figure}[h]
	\centering
	\includegraphics[width=0.6\hsize]{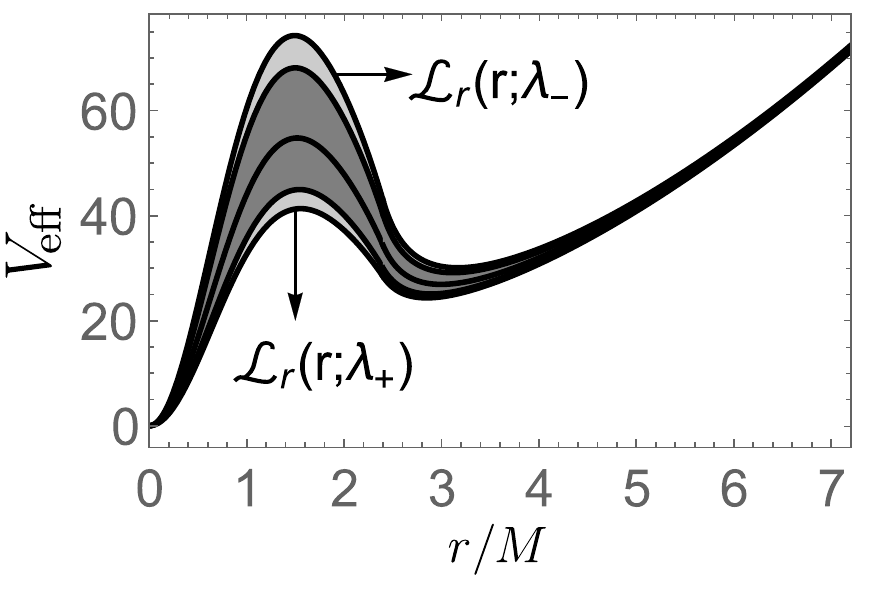}
	\caption{The effective potential
		$\pazocal{L}_r$ for allowed values of the
		parameter $\lambda$. In the shaded area, we find all possible
		effective potentials for the allowed values of lambda and
		$\theta = \pi / 2$. In the darker region, there are effective
		potentials with allowed values $\lambda$ and $\theta = \pi / 4$
		or $3 \pi / 4$, and the curve in the middle of the dark region
		plots the effective potential as $\theta$ approaches
		the poles.}
	\label{f:Vefflim}
\end{figure}

The effective potential in the first-order 
Hartle-Thorne spacetime 
depends on the parameter $\lambda$ that governs the motion in the axial 
direction. At a given radius $r$ and latitude $\theta=\pi/2$ 
corresponding to the equatorial plane of the rotating object, the 
limiting values of $\lambda$ given by the functions 
$\lambda_{r\pm}(r,\theta=\pi/2)$ govern the maximally (exactly) 
co-rotating ($\lambda_{r+}(r,\theta=\pi/2)$) and 
maximally (exactly) counter-rotating 
($\lambda_{r-}(r,\theta=\pi/2)$) null geodesics. The 
effective potentials for those impact parameters (directions) represent 
the boundary of the effective potentials corresponding to the null 
geodesics in all other directions with impact parameters $\lambda \in 
(\lambda_{r-}(r,\theta=\pi/2),\lambda_{r-}(r,\theta=\pi/2))$. 
If we consider latitudes $\theta \neq \pi/2$, the 
values of the limiting impact parameters $\lambda_{r\pm}(r,\theta)$ will 
be governed by the general Eq. (\ref{lamlim}), where 
we can observe the shifting factor $\sin^2\theta$ related to the 
vanishing of the rotational effects on the rotation axis $\theta=0$, 
complexified by the additional term depending on the latitude -- see 
Fig. \ref{f:Vefflim}.

The positions of the maximum and minimum of the effective potential 
$\pazocal{L}_{r}(r,\lambda)$, i.e., the positions $r_\mathrm{c(i)}$ and 
$r_\mathrm{c(e)}$, can be easily seen from the functions 
$\lambda_{c}(r)$ and from Fig. \ref{f:lclthj} (the maximum is always 
below the surface of the object and the minimum is 
always above it). Note that these extrema depend on the parameters 
characterizing the spacetime, but also on the impact parameter $\lambda$ 
(see Fig. \ref{f:Vefflim}), similarly to the values of the radii giving 
the limits of the trapping region: the radius governing 
motion fully constrained to the internal spacetime ($r_\mathrm{c(i)}$ 
that is the solution of the equation $\pazocal{L}_{r}(r,\lambda) = 
\pazocal{L}_{r}(R,\lambda)$) and the radius allowing for motion in the 
external spacetime ($r_\mathrm{c(e)}$ that is the solution of 
$\pazocal{L}_{r}(r,\lambda) = \pazocal{L}_r(r_\mathrm{c(e)},\lambda)$). These 
complexities have to be considered in calculating the coefficient 
characterizing the trapping effect globally for the whole of the 
first-order Hartle-Thorne spacetime.

\begin{figure}[h]
	\centering
	\includegraphics[width=0.8\hsize]{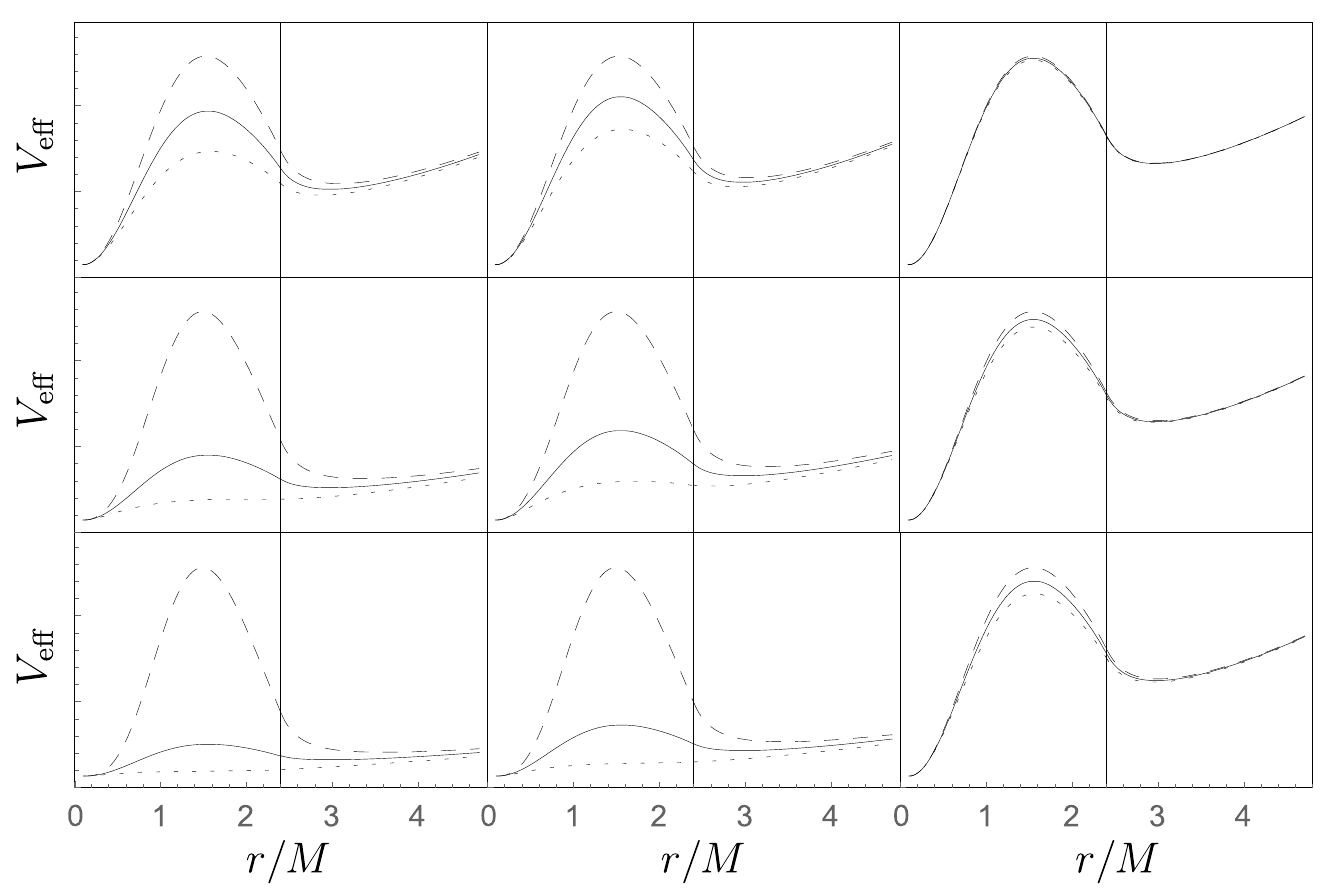}
	\caption{Effective potentials for objects
		with compactness $R/M=2.4$ for several values of $j$ and
		$\theta$. The solid lines are effective
		potentials for non-rotating configurations.
		The dotted and dashed lines denote
		effective potentials for co-rotating and
		counter-rotating null geodesics 
		($\lambda=\lambda_+$ and
		$\lambda=\lambda_-$ respectively). The first column is for
		$\theta=\pi/2$, the second is for $\theta=\pi/4$ and
		the last is for $\theta=1/100$. The
		first row is for $j=0.1$, the second is for
		$j=0.4$ and the last is for $j=0.7$.}
	\label{f:rm24}
\end{figure}

\begin{figure}[h]
	\centering
	\includegraphics[width=0.8\hsize]{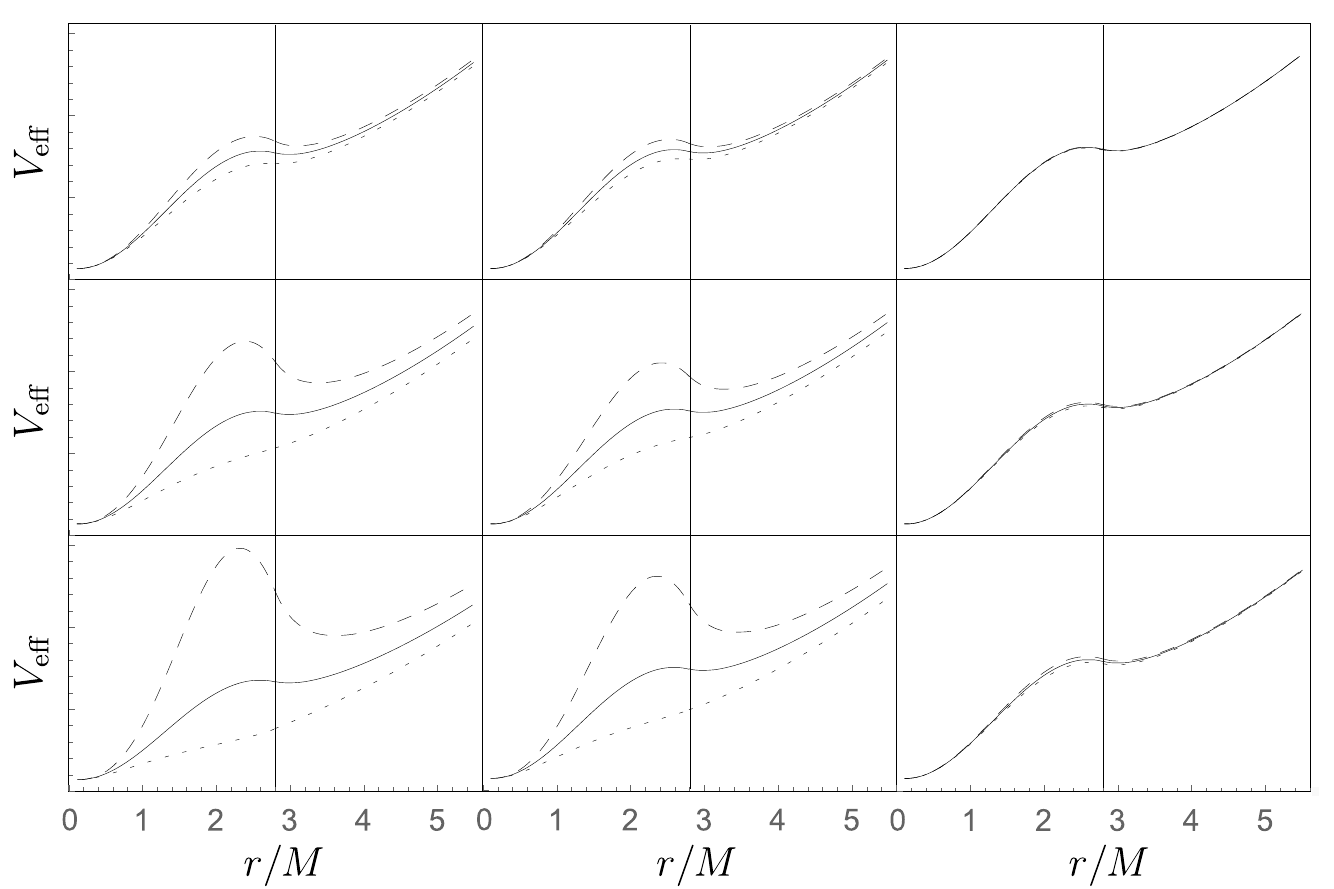}
	\caption{Effective potentials for objects
		with compactness $R/M=2.8$ for several values of $j$ and
		$\theta$. The solid lines are effective
		potentials for non-rotating configurations.
		The dotted and dashed lines denote
		effective potentials for co-rotating and
		counter-rotating null geodesics 
		($\lambda=\lambda_+$ and
                $\lambda=\lambda_-$ respectively). The first column is for
		$\theta=\pi/2$, the second is for $\theta=\pi/4$ and
		the last is for $\theta=1/100$. The
		first row is for $j=0.1$, the second is for
		$j=0.4$ and the last is for $j=0.7$.}
	\label{f:rm28}
\end{figure}

\begin{figure}[h]
	\centering
	\includegraphics[width=0.8\hsize]{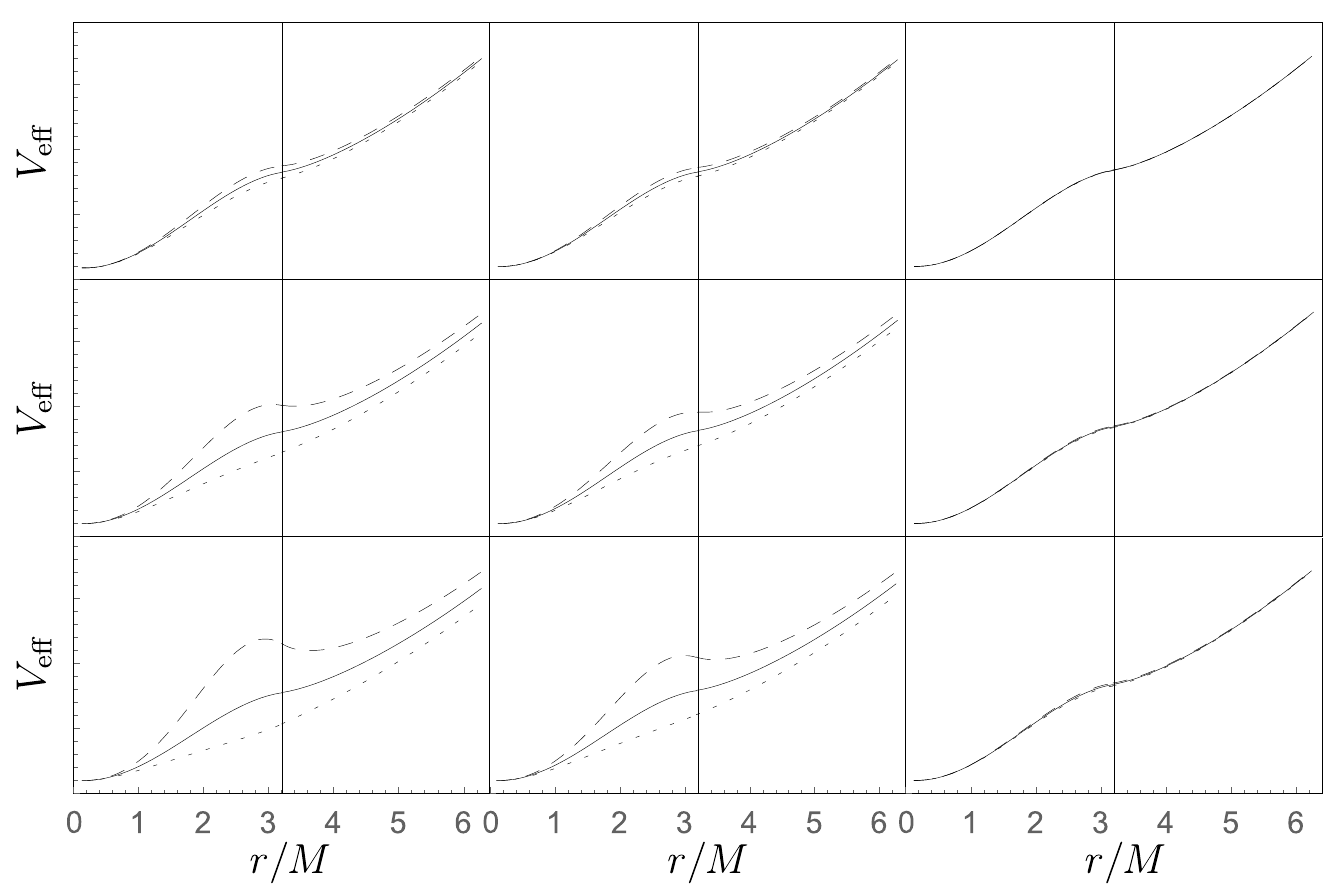}
	\caption{Effective potentials for objects
		with compactness $R/M=3.2$ for several values of $j$ and
		$\theta$. The solid lines are effective
		potentials for non-rotating configurations.
		The dotted and dashed lines denote
		effective potentials for co-rotating and
		counter-rotating null geodesics 
		($\lambda=\lambda_+$ and
                $\lambda=\lambda_-$ respectively). The first column is for
		$\theta=\pi/2$, the second is for $\theta=\pi/4$ and
		the last is for $\theta=1/100$. The
		first row is for $j=0.1$, the second is for
		$j=0.4$ and the last is for $j=0.7$.}
	\label{f:rm32}
\end{figure}

The effective potentials related to the maximally 
co-rotating and maximally counter-rotating 
null geodesics (corresponding to the limiting 
values of the impact parameter $\lambda$) are given in Figs. 
\ref{f:rm24} - \ref{f:rm32} for various values of the rotation rate 
represented by the parameter $j$, the spacetime inverse compactness 
$R/M$, and characteristic latitude $\theta$; their comparison with the 
non-rotating effective potential is also given. Note how the effective 
potential near the poles approaches the effective potential related to 
the non-rotating internal Schwarzschild spacetime due 
to the fact that $g_{t\varphi}$ goes to zero at the poles \footnote{Of 
course rotation represented here by $\bar\omega$ (with respect to the 
local non-rotating frames), or rotation of the 
spacetime (frame dragging), given by $\omega$ are not 
zero at the poles if $\Omega$ is non-zero.}.  We 
include also relatively high values of the rotational 
parameter $j$ in order to stress and clearly illustrate the role of the 
rotation in the trapping effects. We can see that increasing rotation 
causes increase of trapping of counter-rotating null 
geodesics, while it 
decreases trapping of co-rotating null geodesics. 
This phenomenon is enhanced with increasing compactness.

The monotonicity of the effective potential in the internal
Schwarzschild spacetimes with $R/M>3$ implies that rotating
configurations with inverse compactness $R/M>3$ have a monotonic
effective potential near the poles (independent of
the rotation rate), see
Fig. \ref{f:rm32}. In the same figure, we can also
see that for inverse compactness $R/M=3.2$ and a small rotation rate
there is no minimum and maximum of the effective potential, and there is
no trapping of null geodesics as already mentioned in the discussion of
Fig. \ref{f:lclthj}. However, with high enough rotation
parameter, $j > 0.3$, trapping of counter-rotating 
null geodesics is possible, but the co-rotating null 
geodesics cannot be trapped.

\begin{figure}[h]
	\centering
	\includegraphics[width=1.\hsize]{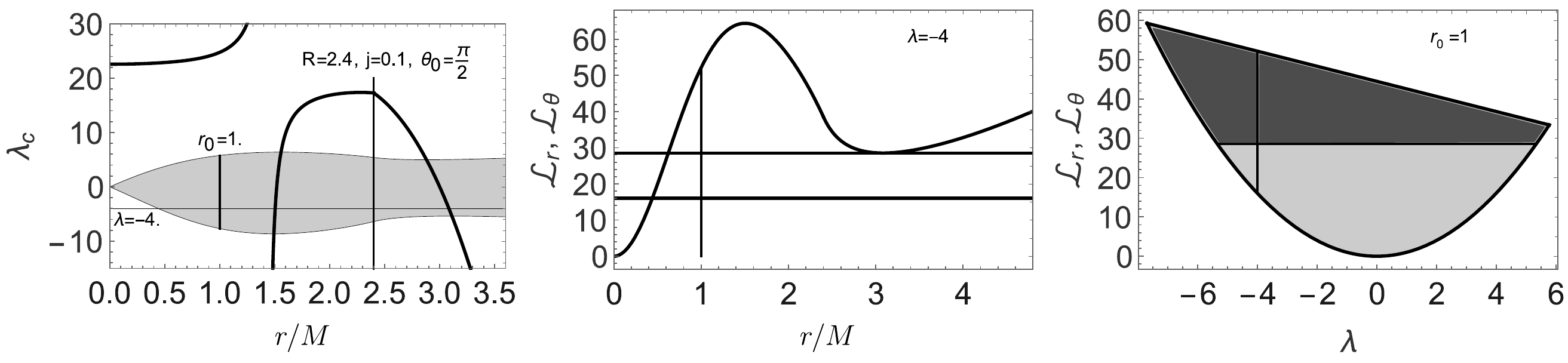}
	\caption{The left panel shows the 
		restriction on the values of $\lambda$ and the function 
		$\lambda_c$. The middle panel shows the effective potentials 
		$\pazocal{L}_r$ and $\pazocal{L}_\theta$, and the minimum of 
		$\pazocal{L}_r$; the right panel shows the allowed values of 
		$\pazocal{L}$ with their dependence on $\lambda$ (shaded areas, 
		with the trapped region being darker). All of the figures are 
		for the specific case of null geodesics emitted from the 
		equatorial plane at radius $r_0=1$ with $\lambda=-4$.}
	\label{f:Lllc}
\end{figure}

For the construction of the trapping cones (complementary escape cones)
Fig. \ref{f:Lllc} is highly instructive and
illustrative, as it demonstrates how the characteristic functions of
$\lambda$ give the corresponding effective potentials $\pazocal{L}_r$
that enable determination in a given position
characterized by coordinates $r,\theta$ of the range
of values of the impact parameters $\lambda$ and $\pazocal{L}$
corresponding to the trapped (escaping) null geodesics.

\FloatBarrier
\section{Trapping cones in first-order Hartle-Thorne 
spacetime}
In the internal Schwarzschild spherically symmetric 
spacetimes the trapping (escape) cones are centrally symmetric as they 
are dependent on a single impact parameter due to the symmetry; in the 
axially symmetric Hartle-Thorne internal spacetimes the symmetry is 
naturally broken. In the first-order Hartle-Thorne 
spacetimes they are dependent on two impact parameters.

\subsection{Construction of trapping cones in the first-order Hartle-Thorne spacetime}

The motion along the null geodesic is fully determined by the motion 
constants (impact parameters) $\lambda$ and $\pazocal{L}$ that can be 
related to any pair of the directional angles 
\{$\alpha,~\beta,~\gamma$\} defined as shown in Fig. \ref{f:defangles}. 
The relation between the tetrad components of the wave 
four-vector of the null geodesics and the directional angles related to 
the tetrad is
\begin{eqnarray}
&&k^{(t)}=-k_{(t)}=1,\\
&&k^{(r)}=k_{(r)}=\cos{\alpha},\\
&&k^{(\theta)}=k_{(\theta)}=\sin{\alpha}\cos{\beta},\\
&&k^{(\varphi)}=k_{(\varphi)}=\sin{\alpha}\sin{\beta}=\cos{\gamma}.
\label{e:mom}
\end{eqnarray}
The directional angle $\alpha$ is determied by the relation 
\begin{eqnarray}
\cos{\alpha}=\frac{k^{(r)}}{k^{(t)}}
=\frac{e^{(r)}_{\mu}k^{\mu}}{e^{(t)}_{\mu}k^{\mu}}.
\label{e:cosa}
\end{eqnarray}
Using Eq. (\ref{e:mom}) in Eq. (\ref{e:constlal}), it turns out that the 
impact parameter $\lambda$ depends only on the angle $\gamma$
\begin{eqnarray}
\lambda=\frac{e_{\varphi}^{(\mu)}k_{(\mu)}}
{-e_{t}^{(\mu)}k_{(\mu)}}
=\frac{e_{\varphi}^{(t)}k_{(t)}
+e_{\varphi}^{(\varphi)}k_{(\varphi)}}
{-e_{t}^{(t)}k_{(t)}-e_{t}^{(\varphi)}k_{(\varphi)}}
=\frac{-e_{\varphi}^{(t)}+e_{\varphi}^{(\varphi)}\cos{\gamma}}
{e_{t}^{(t)}-e_{t}^{(\varphi)}\cos{\gamma}}.
\label{e:lamgam}
\end{eqnarray}

In determining the trapping (escaping) cones we 
strictly follow the method developed and applied in 
\cite{Schee09,Stuchlik10,Stuchlik18} where all details of our approach can be 
found. By setting $\gamma$ and a given point of the selected 
first-order 
Hartle-Thorne spacetime, characterized by coordinates ($r,\theta$), we 
can determine the impact parameter $\lambda$ and the related effective 
potential $\pazocal{L}_{r}(r,\lambda)$ giving limits on the impact 
parameter $\pazocal{L}$. To find out if a photon is trapped (or escapes 
to infinity) and thus construct the trapping (escape) cone, we have to 
know the minimum of $\pazocal{L}_r$ coming from d$\pazocal{L}_r/$d$r$=0. 
From the intersection of $\pazocal{L}(r_\mathrm{c(e)},\lambda)$ 
($\pazocal{L}_r$ at the minimum) and 
$\pazocal{L}_\theta(\lambda)$ (which defines minimal allowed value of 
$\pazocal{L}$), we get a restriction on the values of $\lambda$ and 
thanks to Eq. (\ref{e:lamgam}) we automatically obtain 
the restriction on the directional angle $\gamma$. 
Generally allowed $\lambda$ values can be seen in the 
right-hand panel of Fig. \ref{f:Lllc} (grey areas).

The values of the directional angles $\gamma\in 
\langle\gamma_{min},\gamma_{max}\rangle$ are the cut-off values 
separating photons escaping to infinity and those trapped by gravity. 
Using Eq. (\ref{e:cosa}), we get the limiting value of the directional 
angle $\alpha$ and subsequently from Eq. (\ref{e:mom}) also the limiting 
value of the directional angle $\beta$. The trapping cone is formed by 
plotting all of the limiting angles [$\alpha, \beta$]. Note that the region 
of co-rotating null geodesics contains the angle 
$\beta=\pi/2$, while the region of counter-rotating 
null geodesics contains $\beta=3\pi/2$; the separation line is 
given by the angles $\beta=0$ and $\beta=\pi$.

\begin{figure}[h]
	\centering
	\includegraphics[width=0.90\hsize]{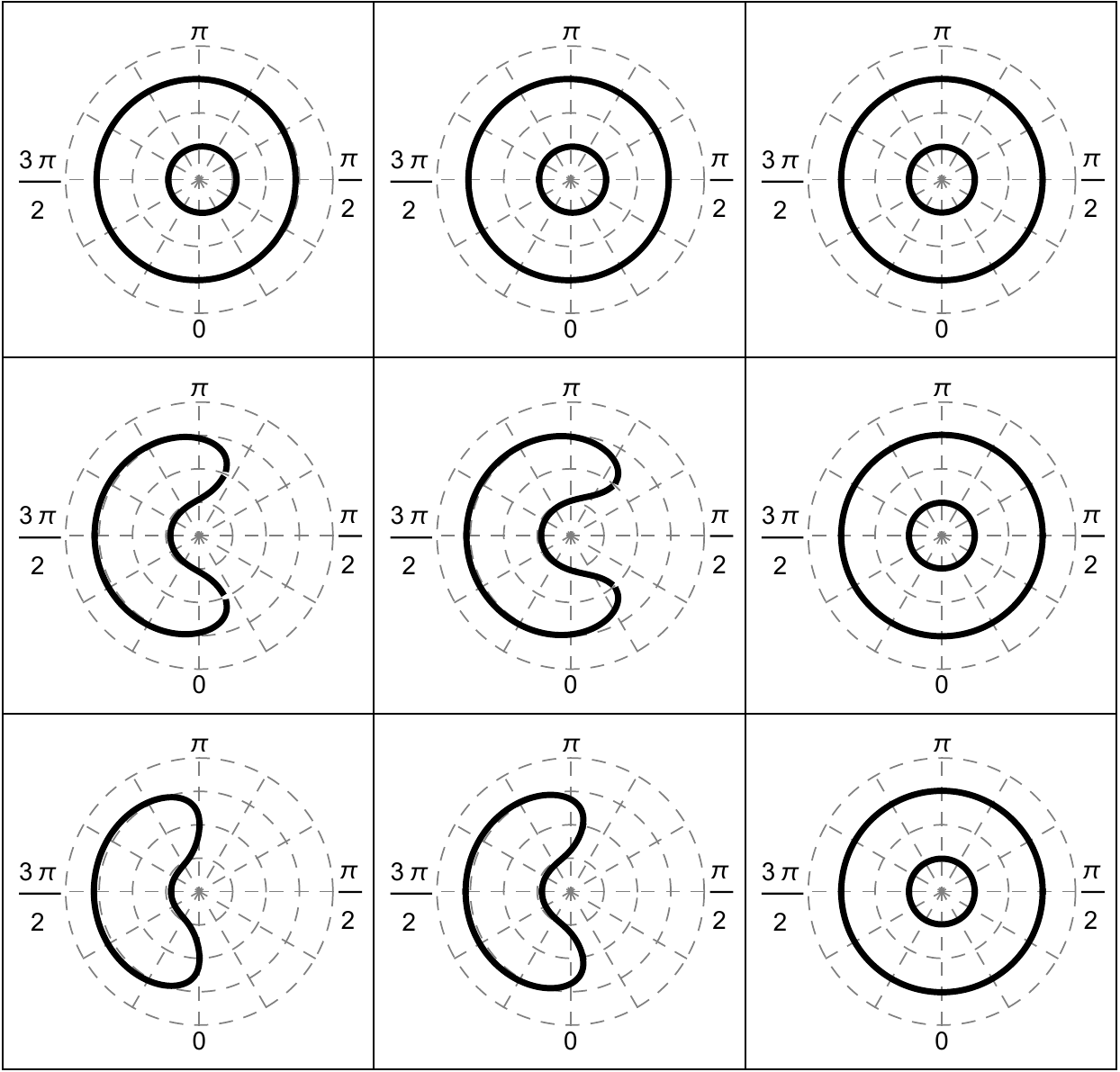}
	\caption{Escape cones for objects with 
		compactness $R/M=2.4$ for 
		several values of $\theta$ and $j$. The first column is for 
		$\theta=\pi/2$, the second column is for $\theta=\pi/4$ and 
		the last column is for $\theta=1/100$. The 
		first row is for $j=0.1$, the second row is 
		for $j=0.4$ and the last row is for 
		$j=0.7$.}
	\label{krm24}
\end{figure}

\begin{figure}[h]
	\centering
	\includegraphics[width=0.90\hsize]{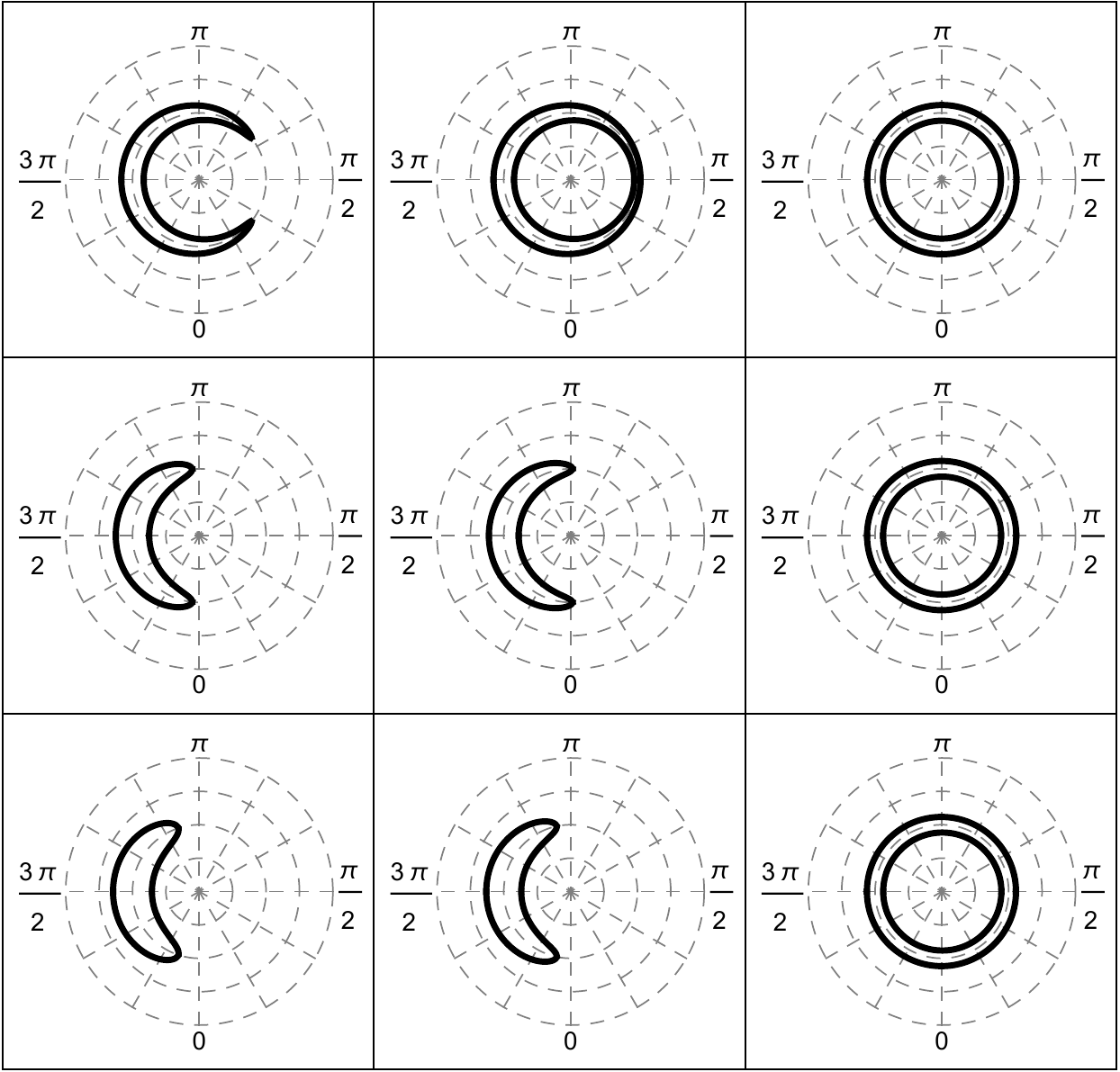}
	\caption{Escape cones for objects with 
		compactness $R/M=2.8$ for 
		several values of $\theta$ and $j$. The first column is for 
		$\theta=\pi/2$, the second column is for $\theta=\pi/4$ and 
		the last column is for $\theta=1/100$. The 
		first row is for $j=0.1$, the second row is 
		for $j=0.4$ and the last row is for 
		$j=0.7$.}
	\label{krm28}
\end{figure}

\begin{figure}[h]
	\centering
	\includegraphics[width=0.90\hsize]{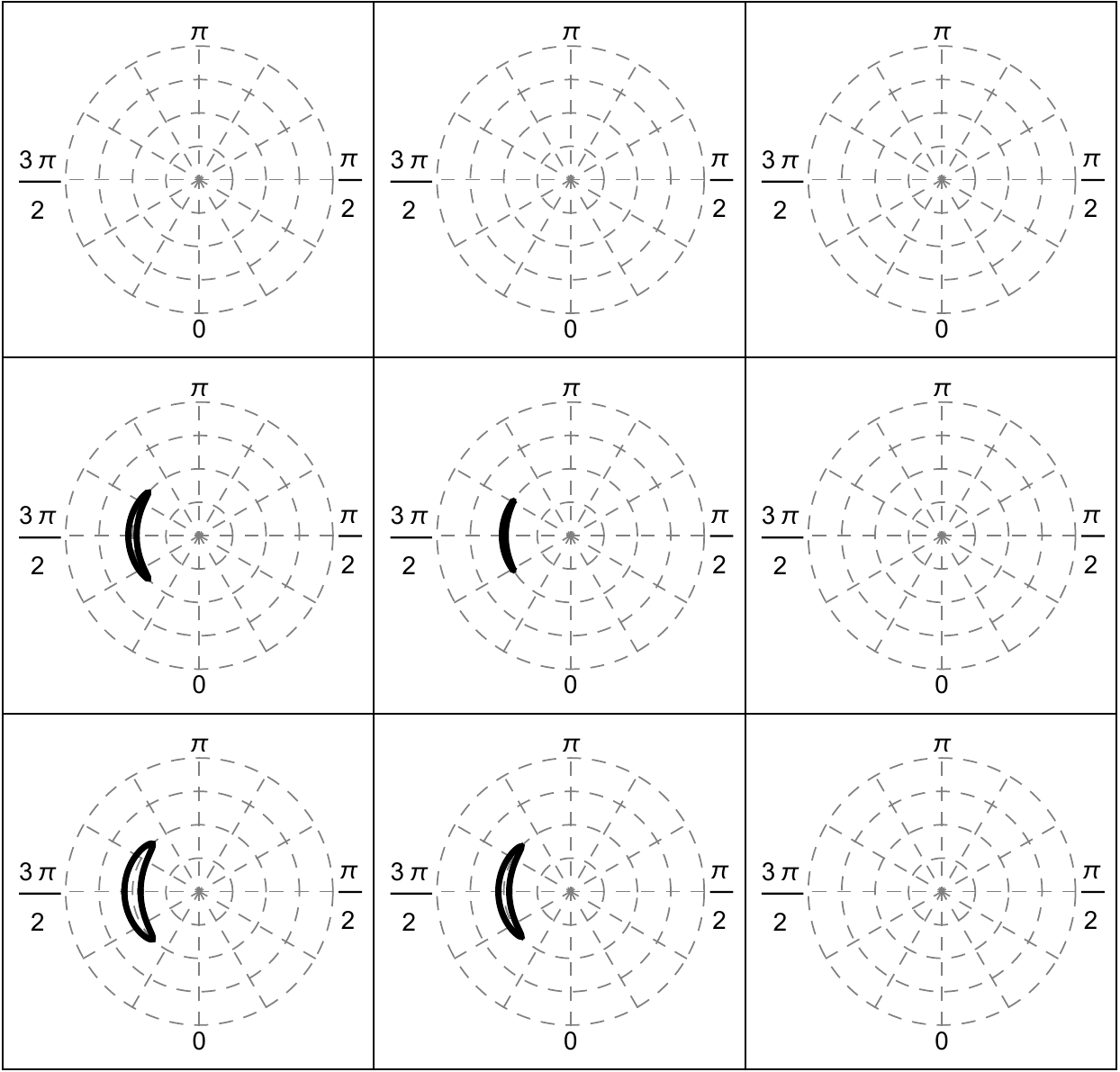}
	\caption{Escape cones for objects with 
		compactness $R/M=3.2$ for 
		several values of $\theta$ and $j$. The first column is for 
		$\theta=\pi/2$, the second column is for $\theta=\pi/4$ and 
		the last column is for $\theta=1/100$. The 
		first row is for $j=0.1$, the second row is 
		for $j=0.4$ and the last row is for 
		$j=0.7$.}
	\label{krm32}
\end{figure}

The resulting Figs. \ref{krm24} - \ref{krm32} are presented for the same 
selection of the spacetime parameters and the latitudes of the cone 
position, with the radial coordinate being fixed at $r/M=1$ as in the 
case of the representative figures for the effective potentials. These 
figures clearly demonstrate the effects described above, especially the 
breaking of the central symmetry that increases with increasing values 
of $R/M$ and, of course, it decreases with decreasing 
values of the latitudinal coordinate of the position 
of the trapping cone as the influence of rotation effects is weakened as 
the apex of the cone approaches the symmetry axis.

The profiles of the effective potentials 
(co-rotating and counter-rotating) 
with small rotation rates and $R/M=3.2$ imply that the 
trapping cone can exist only in a small region of 
counter-rotating null geodesics and only in 
spacetimes with a sufficiently high rotation parameter $j$.

\FloatBarrier
\section{Null geodesics trapped in the slowly rotating spacetime}

Finally, we are able to study the trapping of neutrinos and the local 
and global trapping efficiency as they depend on 
the spacetime parameters and, in the case of the 
	local trapping, also as they depend on the position of the radiation 
source. We always assume a locally isotropic source 
of the neutrinos following the null geodesics, as discussed in 
\cite{Stuchlik09b}.

\subsection{Local trapping}

We vary $\theta$, $j$ and $R/M$ in the same way as 
in the presentation of the representative cases for 
the effective potential of null geodesic motion 
$\pazocal{L}_{r}(r,\lambda)$ and the trapping (escape) cones, and we 
give the radial profiles of the local trapping efficiency 
at fixed latitudinal coordinates of the compact 
object, considering the region allowing for the trapping. We follow the 
methodology introduced in \cite{Stuchlik09b}, including the assumption 
of the isotropically radiating sources.

In order to reflect the local properties of the trapping effect we 
introduce a local trapping efficiency coefficient $b$, defined at a 
given point of the compact object determined by the 
coordinates $r,\theta$, as the ratio of the number 
of neutrinos emitted from this point and trapped by 
the object $N_b$ to the total number of neutrinos 
produced at this point $N_p$ (for details see \cite{Stuchlik09b}). Due 
to the isotropy of the radiation emitted by the local source, the local 
trapping coefficient is given by the ratio of the surface 
area of the trapping cone $S_{\mathrm{tr}}$ to the 
total area $S_\mathrm{tot}=4\pi$. Therefore,

\begin{equation}
b(r,\theta;j,R)\equiv\frac{\mathrm{N_b(r)}}{\mathrm{N_p(r)}} 
= \frac{S_{\mathrm{tr}}}{4\pi}. 
\end{equation}

\begin{figure}[h]
	\centering
	\includegraphics[width=0.90\hsize]{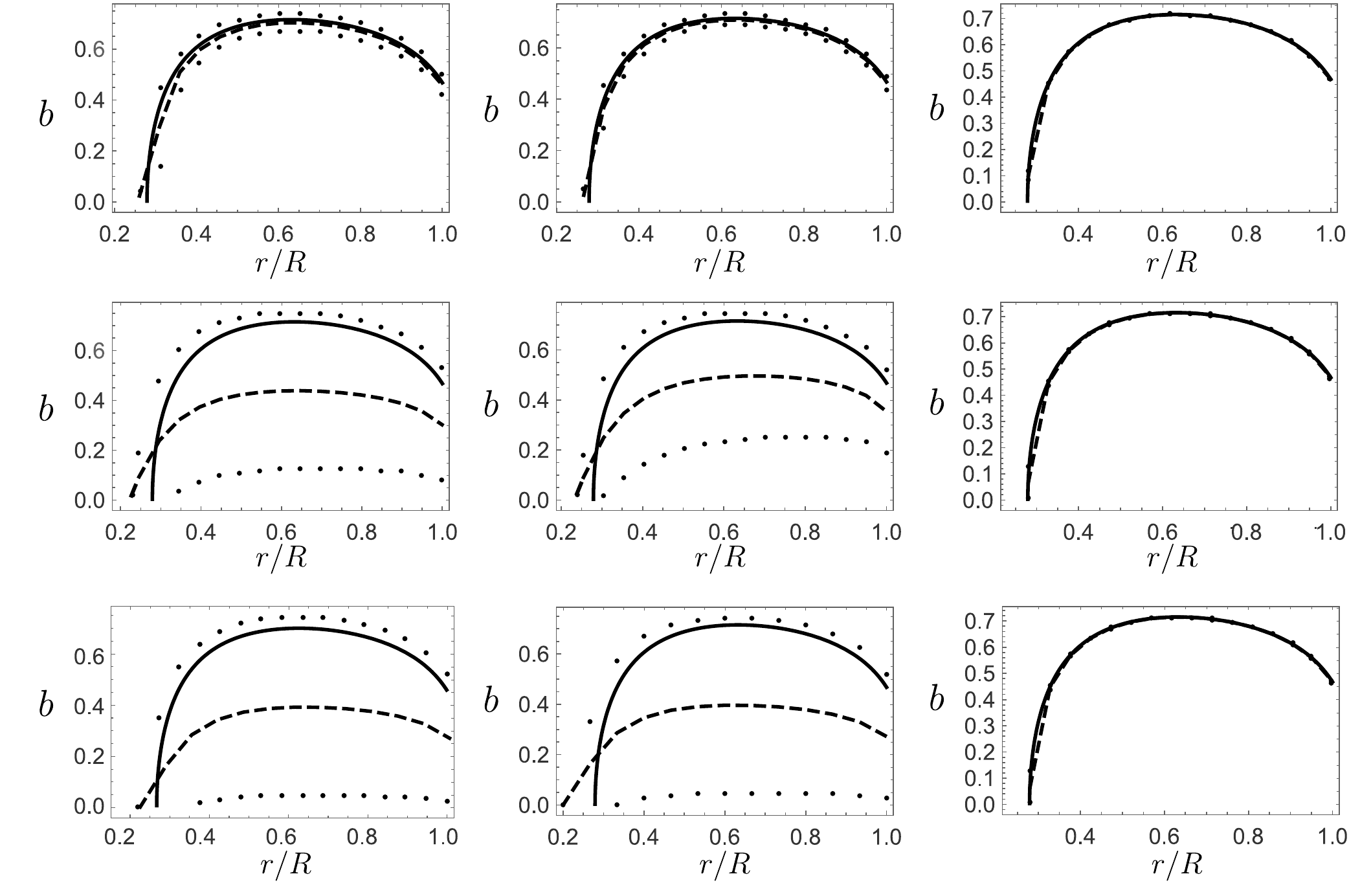}
	\caption{The local trapping efficiency coefficient $b$ for 
		$R/M=2.4$ is plotted for several values of $j$ and $\theta$. 
		The solid line shows the local trapping for a non-rotating 
		configuration and the dashed line shows local trapping 
		for the rotating case. For the latter, 
		the points above the dashed line are values for just 
		counter-rotating directions of the null 
		geodesics, and the points below it are those for 
		co-rotating directions. The first column 
		is for $\theta=\pi/2$, the second is for $\theta=\pi/4$ and 
		the last is for $\theta=1/100$. The first row is for $j=0.1$, 
		the second is for $j=0.4$ and the last is for $j=0.7$.}
\label{loct24}
\end{figure}

\begin{figure}[h]
\centering
\includegraphics[width=1\hsize]{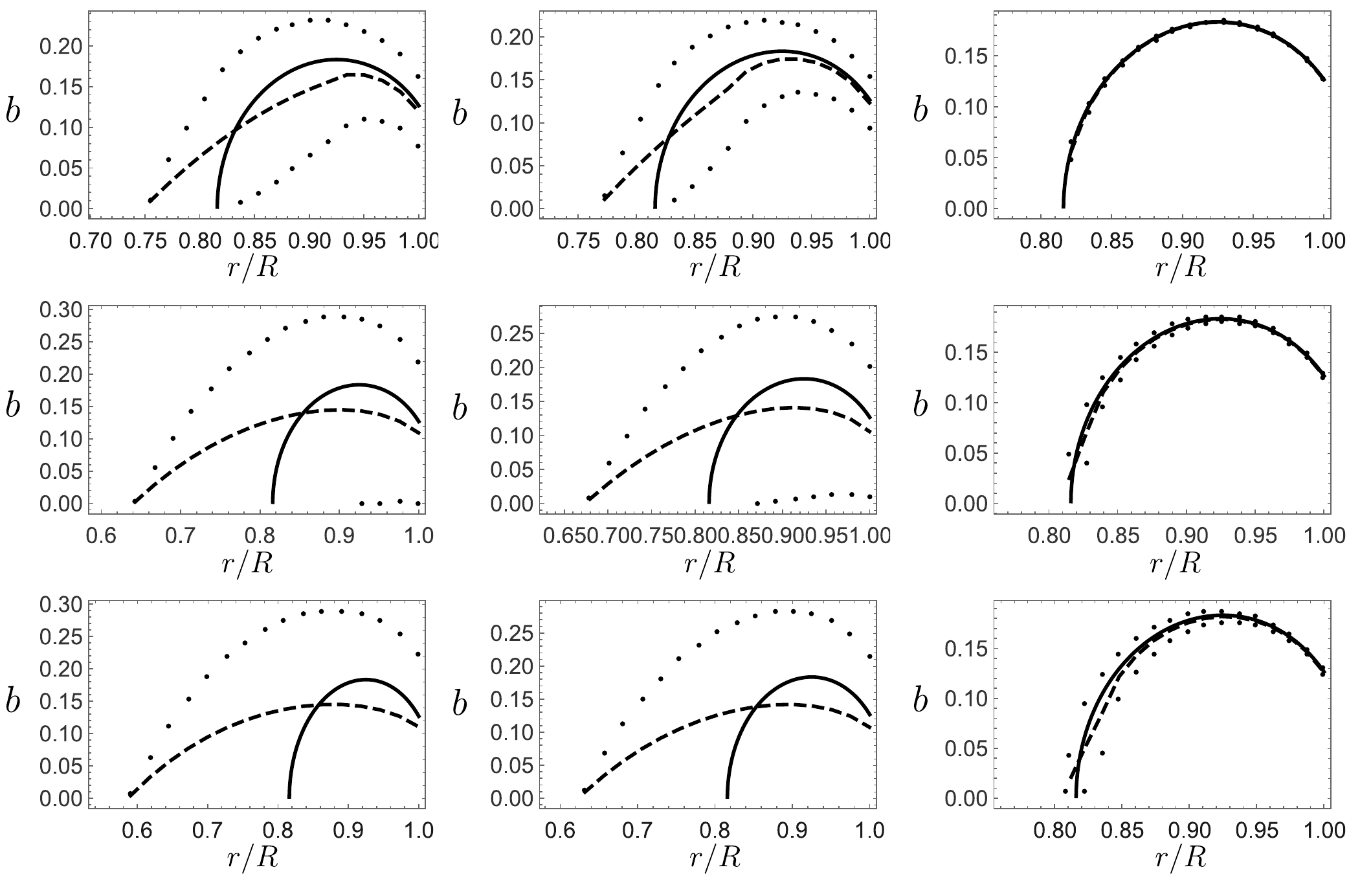}
\caption{The local trapping efficiency coefficient $b$ for 
$R/M=2.8$ is plotted for several values of 
$j$ and $\theta$. The solid line shows the 
local trapping for a non-rotating configuration 
and the dashed line shows local trapping 
for the rotating case. For the latter, 
the points above the dashed line are values for just 
counter-rotating directions of the null geodesics, and the 
points below it are those for co-rotating directions. 
The first column is for $\theta=\pi/2$, the 
second is for $\theta=\pi/4$ and the last 
is for $\theta=1/100$. The first row is for $j=0.1$, the second
is for $j=0.4$ and the last 
is for $j=0.7$.}
\label{loct28}
\end{figure}

\begin{figure}[h]
\centering
\includegraphics[width=1\hsize]{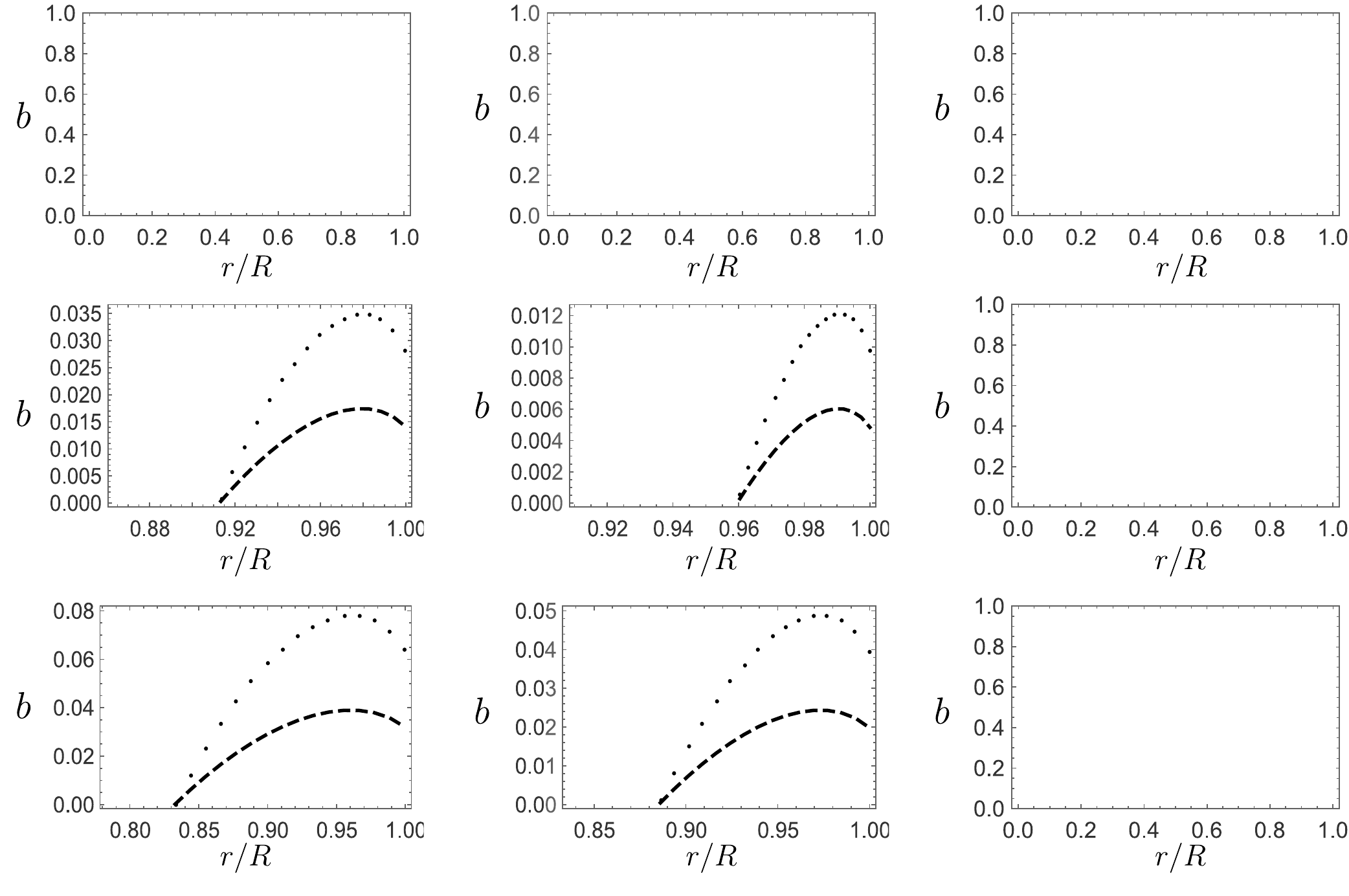}
\caption{The local trapping efficiency coefficient $b$ for 
$R/M=3.2$ is plotted for several values of 
$j$ and $\theta$. 
There is no trapping here for the non-rotating 
configuration and trapping only for counter-rotating directions in
the rotating case. The local trapping for the rotating case is again 
shown with the dashed lines and the points above them show results
for just the counter-rotating directions of the null geodesics 
(giving twice the values for the dashed curves).
The first column is for $\theta=\pi/2$, the 
second is for $\theta=\pi/4$ and the last 
is for $\theta=1/100$. The first row is for $j=0.1$, the second
is for $j=0.4$ and the last 
is for $j=0.7$.}
\label{loct32}
\end{figure}

We use this procedure for all radii $r$ relevant for the trapping, i.e., 
$r \geq r_\mathrm{b(e)}$, see Fig. \ref{svef}, where 
$r_\mathrm{b(e)}(\theta,R/M,j)$ is the radius where the trapping begins. 
The resulting local trapping efficiency coefficient 
is presented in Figs. \ref{loct24} - \ref{loct32}. For comparison and 
better insight into the problem, we depict also the analytical solution 
obtained for the internal Schwarzschild spacetime in \cite{Stuchlik09b} 
(solid line). In order to clearly illustrate the role of the rotation of 
the object, we calculate separately the trapping 
coefficients for the co-rotating and 
counter-rotating neutrinos, thus splitting the 
coefficient $b$ into two complementary parts denoted as $b_+$ and $b_-$; 
$b_+$ denotes the local trapping efficiency coefficient for 
co-rotating 
null geodesics represented by the right part of the trapping cones 
(the part containing $\beta=\pi/2$), while $b_-$ 
denotes the coefficient calculated for counter-rotating 
null geodesics corresponding to the left part of the trapping cones, 
containing the angle $\beta=3\pi/2$. We thus define
\begin{equation}
b_{+}(r,\theta;j,R) = \frac{S_{\mathrm{tr}_+}}{2\pi} \textrm{, } 
b_{-}(r,\theta;j,R) = \frac{S_{\mathrm{tr}_-}}{2\pi} .
\end{equation}

In our figures, $b_-$ tends to be located above the analytical function 
representing the case of the spherically symmetric spacetime (if it 
exists) and $b_+$ tends to be located below this function.

We can see that generally the local trapping in the rotating 
Hartle-Thorne spacetime is lower than that for the 
internal Schwarzschild spacetimes, with the 
exception of the deepest regions of the trapping that usually reach 
smaller radii in the rotating spacetimes (especially for the 
counter-rotating null geodesics). The extension of the 
trapping region in the rotating spacetimes, in comparison with the 
related internal Schwarzschild spacetime increases with increasing 
parameter $R/M$. As expected, in the first-order 
Hartle-Thorne 
spacetimes with $R/M > 3$, the trapping effect can be relevant 
only for the counter-rotating null geodesics, and 
we observe the trapping only for sufficiently high rotation parameters, 
$j > 0.2$.

\subsection{Global trapping}

The coefficient of global trapping reflects the trapping phenomenon 
integrated across the whole trapping region, related to the whole 
radiating object. We thus consider the number of neutrinos radiated 
along null geodesics by the whole object in unit time for distant static 
observers, and determine the fraction of these 
radiated neutrinos that remain trapped by the radiating object. Details 
of the derivation of the global trapping coefficient are presented in 
\cite{Stuchlik09b}, and we apply them here using again 
the basic assumption that the locally defined radiation intensity is 
proportional to the energy density of the matter of the object 
hence being distributed uniformly within all of the 
object.

The global trapping effects are then reflected by the 
global trapping efficiency coefficient $\pazocal{B}$ defined by the relation 
\cite{Stuchlik09b}
\begin{equation}
\pazocal{B}\equiv\frac{N_b}{N_p}
=\frac{\int_{0}^{R}\int^{\pi}_{0}\int^{2\pi}_{0}f^{-1}(r)
	b(r) \, r^2 \, \mathrm{d}\varphi \,\mathrm{d}\theta \, \mathrm{d}r}
{\int_{0}^{R}\int_{0}^{\pi}\int_{0}^{2\pi}f^{-1}(r) 
	\, r^2 \, \mathrm{d}\varphi \,\mathrm{d}\theta \, \mathrm{d}r},
\end{equation}
where we use the radial metric function $f(r)$ defined in Eq. (\ref{e:f}).

However, we can simplify the integration process 
due to the symmetries of the first-order 
Hartle-Thorne spacetime. The 
metric coefficients are independent of $\varphi$, and the results of 
the integration in the upper 
and lower hemispheres are the same. Also, limits in the radial 
direction can be shrunk by using knowledge of the position of 
$r_\mathrm{b(e)}(r,\theta;R,J,\lambda)$ determining the limits of integration 
of the trapping effect. The global trapping 
efficiency coefficient can then be presented in the 
form
\begin{equation}
\pazocal{B}=\frac{\int_{r_\mathrm{b(i)}}^{R}\int^{\pi/2}_{0}
f^{-1}(r)b(r) \, r^2 \, \mathrm{d}\theta \, \mathrm{d}r}
{\int_{0}^{R}\int_{0}^{\pi/2}f^{-1}(r) \, r^2 \, 
\mathrm{d}\theta \, \mathrm{d}r}.
\end{equation}
	
\begin{figure}[h]
	\centering
	\includegraphics[width=1.05\hsize]{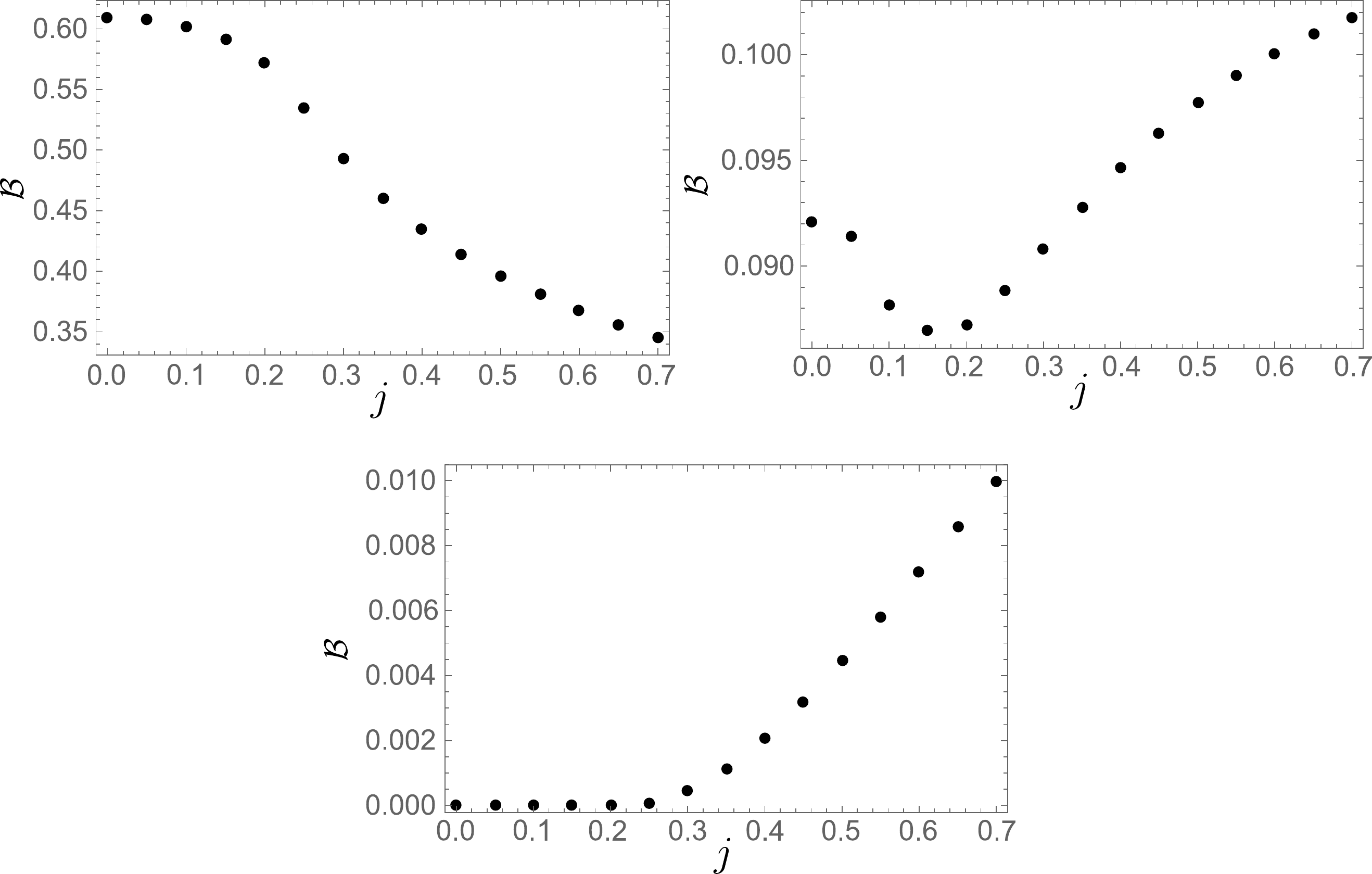}
	\caption{The global trapping efficiency coefficient 
		$\pazocal{B}$ for $R/M=2.4$ is shown in the 
		top left panel, with that for $R/M=2.8$ in the top right panel 
		and that for $R/M=3.2$ in the panel below.}
	\label{globtv}
\end{figure}

\begin{figure}[h]
	\centering
	\includegraphics[width=0.7\hsize]{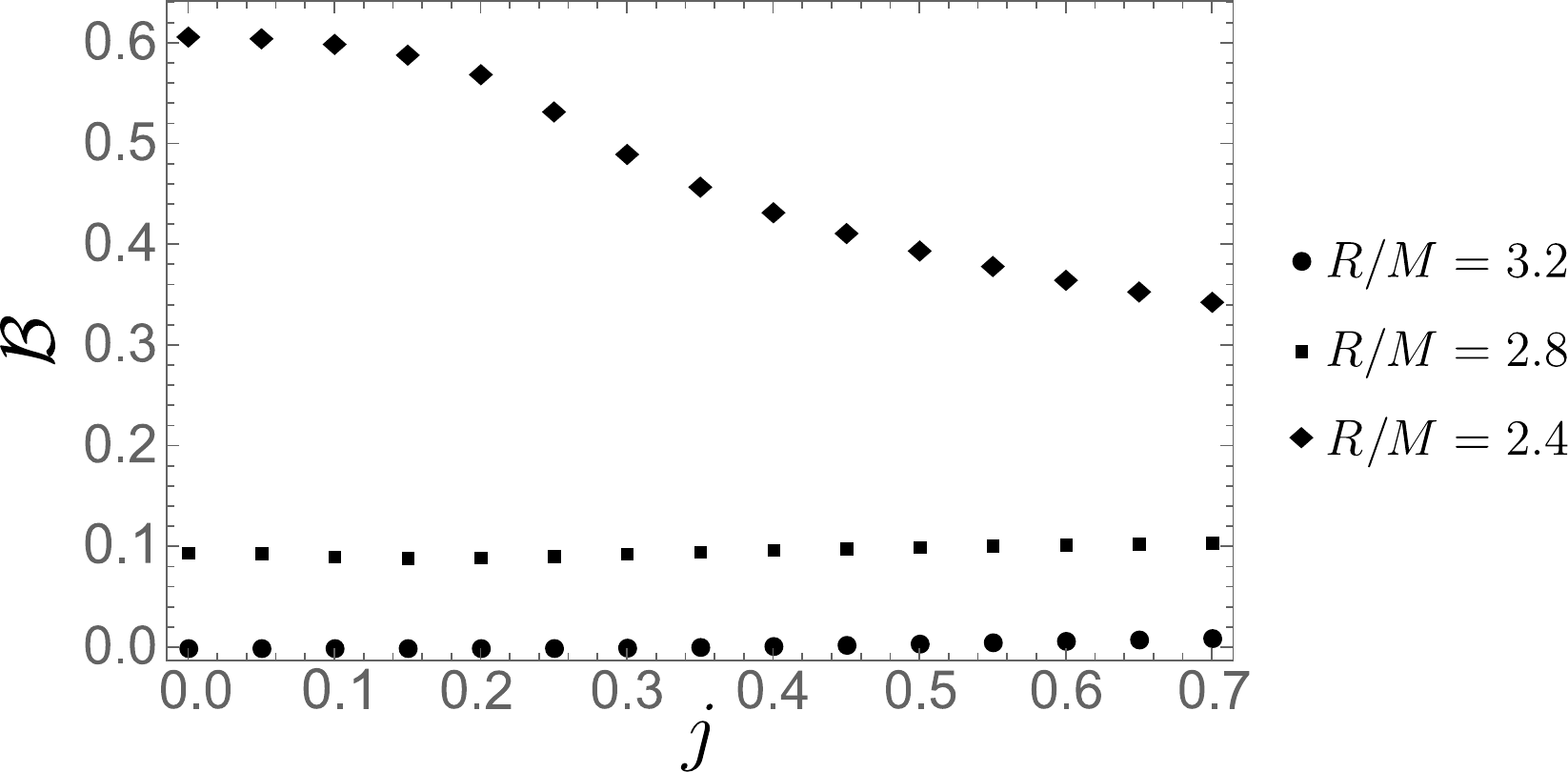}
	\caption{The comparison of the global trapping efficiency 
		coefficients $\pazocal{B}$ from 
		Fig. \ref{globtv} with the curves being shown 
		together with the same vertical scale.}
	\label{glob1}
\end{figure}

Figs. \ref{globtv} and \ref{glob1} show how the global trapping 
efficiency coefficient $\pazocal{B}$ depends on the rotation parameter 
$j$ for fixed characteristic values of the inverse compactness $R/M$ of 
the object\footnote{In order to indicate the possible behavior of the 
trapping effects for the standard internal Hartle-Thorne spacetimes, we 
consider here also rather large values of the rotation 
parameter $j$, up to $j=0.7$, exceeding the value $j=0.5$ which we would 
normally consider as the maximum acceptable for using with the second-order 
Hartle-Thorne approximation} \cite{Urbanec13}..

Note that for the lowest value of the inverse compactness parameter, 
namely $R/M=2.4$, the global efficiency parameter $\pazocal{B}(j)$ 
decreases monotonically with increasing $j$, while for the 
middle value, $R/M=2.8$, we observe a decrease of 
$\pazocal{B}$ up to $j \sim 0.2$ and then a rapid increase for $j>0.2$. 
For the largest value, $R/M=3.2$, where trapping is impossible in the 
non-rotating internal Schwarzschild spacetimes, the 
trapping effect occurs at $j \sim 0.25$ and increases with increasing 
rotation parameter $j$, but exclusively for the 
counter-rotating null geodesics. Nevertheless, we have 
to say that detailed calculations in the second-order 
Hartle-Thorne geometry would be necessary in order to obtain a realistic 
description of the trapping phenomenon for rotation parameters larger 
than the limit of $j = 0.1$ up to which using the 
first-order treatment seems to be fully justified.

\FloatBarrier
\section{Conclusions}

In this introductory study of the role of rotation 
for the phenomenon of trapping of null geodesics that could be relevant 
for the motion of neutrinos in the interior of 
neutron stars, we have used the strongest 
simplification of taking the first-order
Hartle-Thorne spacetime with a uniformly 
distributed energy density for the matter, in order 
to obtain simple and easily tractable results. We believe that even such 
a simplification enables one to find the basic 
characteristics of the influence of the rotation of radiating compact 
objects on the effect of trapping in their interior. For these purposes 
we have considered also values of the dimensionless 
rotation parameter $j$ exceeding those safe for 
ensuring the applicability of the first-order 
approximation ($j$ no larger than around 
0.2). Nevertheless, we expect that the results obtained for values of 
$j > 0.2$ could indicate relevant signatures of realistic effects.

Our results related to the local effects indicate a 
much stronger trapping coefficient for the 
counter-rotating null geodesics than for the 
co-rotating ones and even for the internal 
Schwarzschild spacetimes with the same parameter $R/M$. In the rotating 
Hartle-Thorne spacetimes the region of trapping is always larger than in 
the related non-rotating internal Schwarzschild 
spacetimes having the same parameter $R/M$, and 
this difference increases with increasing $R/M$.

Trapping of counter-rotating null geodesics has been 
found even for $R/M>3$, i.e., for values where trapping does not occur 
in the internal Schwarzschild spacetimes. It has been shown to be 
relevant even for objects with $R/M=3.2$, but only for rotation 
parameters starting at $j \sim 0.25$ for which the 
validity of using the first-order form of the Hartle-Thorne metric is 
not secure. Therefore, more detailed models based on the 
second-order Hartle-Thorne geometry are necessary in 
order to confirm the possibility of surpassing the 
limit for trapping effects at the radius $R=3M$, and to follow the 
occurrence of the trapping for values of the rotation parameter 
$j\sim0.2$, where some relevance of the first-order 
approximation is expected, up to the limiting value of $j\sim0.5$ 
appropriate for when the second-order Hartle-Thorne 
metric is used \cite{Urbanec13}.

Our study is related to the models of very compact neutron stars (or 
quark stars) as we are limiting ourselves to internal spacetimes with 
$R/M\leq3.2$. Surprisingly, it has recently been shown that trapping 
polytropic spheres can exist with very large extension 
$R\gg r_g$ - such structures can model dark matter halos related even to 
galaxy clusters\footnote{Then the cosmological constant can put natural 
limits on the extension of the trapping polytropes \cite{Stuchlik16a}.}. 
The trapping polytropes can exist for polytropic index 
$n>2.2$ as shown and discussed in 
\cite{Novotny17,Hod18,Hod18a,Hod18b,Peng18}. The trapping polytropes 
with $n>3.3$ can be extended to large radius ($R\sim 100$ kpc) and large 
mass ($M\sim10^{12}M_\odot$) corresponding to the 
extension and mass of large galaxies, while the 
extension of the trapping zone $R_\mathrm{tr}$ is of 
order $r_g$, and its mass $M_\mathrm{tr}=M(R_\mathrm{tr})$ 
is of order $10^9M_\odot$; then gravitational 
instability of the trapping zone might induce its 
gravitational collapse and creation of a supermassive black hole having 
$M\sim10^9M_\odot$, thus giving a natural explanation 
for the supermassive black holes observed in quasars 
at cosmological redshifts $z\geq6$ (for details see 
\cite{Stuchlik17})\footnote{Stability of polytropic spheres against 
radial pulsations was studied in \cite{Hladik20}, 
with inclusion of influence of the cosmological constant in 
\cite{Posada20}  -- instability of extremely compact 
polytropes was indicated. Instability of perturbative fields in the 
extremely compact regions with trapping of null geodesics located in 
extremely extended polytropes was demonstrated in \cite{Stuchlik17}. 
These results are in agreement with the study of slowly decaying waves in 
general spherically symmetric spacetimes exhibiting trapping of null 
geodesics -- it was demonstrated that the linear waves cannot decay 
faster than logarithmically, suggesting thus nonlinear instability \cite{Keir14}.}.

Generalization of such spherical trapping polytropes to spacetimes with 
rotation considered to the first-order level could be of high interest in 
the framework of models of dark matter halos. This is going to be the 
scope of further studies.
	
\section*{acknowledgements}
	J.V., M.U., and Z.S. acknowledge the institutional support of the 
	Institute of Physics, Silesian University in Opava. J. V. and M.U. 
	were supported by the Czech Grant No. LTC18058.
	
\section*{References}
	
\bibliographystyle{apsrev4-1}  %% BibTeX style
\bibliography{gravreferences}

\end{document}